\documentclass[journal, 12pt, twocolumn, final]{IEEEtran}
\usepackage{amsmath,amssymb,amstext,amsthm}
\usepackage{MnSymbol}
\usepackage{latexsym,subfigure,algorithm}
\usepackage{algorithmic}
\usepackage{graphicx} 
\usepackage{epstopdf}
\usepackage{epsfig}
\usepackage{psfrag}
\usepackage{ifthen}
\usepackage{color}

\newcommand{\equalalign}{\hspace*{-2mm}&=&\hspace*{-2mm}}
\newcommand{\emptyalign}{&&\hspace*{-2mm}}

\begin{document}
\title{Random Access based Reliable\\ Uplink Communication and Power Transfer\\ using Dynamic Power Splitting}
\author{\IEEEauthorblockN{Steven Kisseleff,~\IEEEmembership{Member,~IEEE}, Symeon Chatzinotas,~\IEEEmembership{Senior Member,~IEEE}}, \\
and Bj\"orn Ottersten,~\IEEEmembership{Fellow,~IEEE}
\thanks{
This paper has been presented in part at IEEE ICC 2019 \cite{kisseleff2019ultrareliable}.\newline
The authors are with the Interdisciplinary Centre for Security, Reliability and Trust (SnT), University of Luxembourg, Luxembourg. E-mail: \{steven.kisseleff, symeon.chatzinotas, bjorn.ottersten\}@uni.lu.\newline
This work was supported by the Luxembourg National Research Fund (FNR) in the framework of the FNR-FNRS bilateral project "InWIP-NET : Integrated Wireless Information and Power Networks".\newline

© 2020 IEEE.  Personal use of this material is permitted.  Permission from IEEE must be obtained for all other uses, in any current or future media, including reprinting/republishing this material for advertising or promotional purposes, creating new collective works, for resale or redistribution to servers or lists, or reuse of any copyrighted component of this work in other works.} 
}

\maketitle
\thispagestyle{empty}
\begin{abstract}
Large communication networks, e.g. Internet of Things (IoT), are known to be vulnerable to co-channel
interference. One possibility to address this issue is the use of orthogonal multiple access (OMA)
techniques. However, due to a potentially very long duty cycle, OMA is not well suited for such schemes.
Instead, random medium access (RMA) appears more promising. An RMA scheme is based on transmission of short data packets with random
scheduling, which is typically unknown to the receiver. The received signal, which consists of the
overlapping packets, can be used for energy harvesting and powering of a relay device. Such an energy
harvesting relay may utilize the energy for further information processing and uplink transmission.
In this paper, we address the design of a simultaneous information and power transfer scheme based
on randomly scheduled packet transmissions and reliable symbol detection. We formulate a
prediction problem with the goal to maximize the harvested power for an RMA scenario. In order to solve this problem, we propose a new prediction method, which shows a
significant performance improvement compared to the straightforward baseline scheme. Furthermore, we investigate the complexity of the proposed method and its vulnerability to imperfect channel state information.
\end{abstract}

\section{Introduction}
\label{sec_1}
\subsection{Motivation}
\IEEEPARstart{O}{ne} of the main challenges in large communication networks (e.g. Internet of Things, IoT) and
telemetry systems (e.g. Low Power Wide Area Networks, LPWANs) is the reliable signal
transmission in the presence of strong co-channel interference \cite{atzori2010the}, \cite{raza2017low}. Typically, orthogonal
multiple access (OMA) techniques can be applied in order to create orthogonal sub-channels
and separate the adjacent transmissions thus minimizing the interference. In case of orthogonal separation of multiple data streams and increasing number of streams, the maximum data rate per stream decreases on average. Furthermore, the decrease of the maximum data rate affects the packet length and the delay between the packets. In this context, we assume that the packet transmissions are aligned with the duty cycle of the respective node. Correspondingly, with increasing number of streams, the duty cycle increases as well. Hence, the resulting duty cycle can be very long in case of a large number of network nodes, which is usually undesirable.
One possibility for reducing the duty cycle is to employ a non-orthogonal multiple
access (NOMA), cf. \cite{saito2013nonorthogonal}, where multiple data streams utilize the same time and frequency
resources. However, this technique is typically limited to only a few parallel data streams, thus
providing limited reduction of the duty cycle compared to traditional OMA. Alternatively, random
medium access (RMA) can be used, which usually has a much shorter duty cycle. For RMA, the
transmissions from individual nodes are not jointly scheduled in time, but occur with a typically
known probability, which can be exploited in order to improve the reliability of signal detection
\cite{lieske2016decoding}, \cite{kisseleff2018optimal}. Hence, a distinct advantage of the RMA is high flexibility of the system. In particular, the nodes can freely choose their duty cycles according to their requirements and power consumption. Furthermore, RMA can be easily adapted in presence
of ad hoc nodes, which can start their transmissions at any time. Although such ad hoc nodes
may need to synchronize their transmissions with respect to carrier frequency and timing offset
according to \cite{mengali1997synchronization}, no joint scheduling of transmissions is required in this case. On the other hand,
RMA provides an additional uncertainty for the symbol detection. Hence, the design of a symbol detection scheme, which guarantees a high reliability of communication, is even more challenging with RMA compared to OMA or NOMA.\\
Since RMA implies discontinuous transmissions via short data packets, in the following we
review the related advancements in this research area. The ultimate performance bounds for the
finite blocklength communication have been derived in \cite{polyanskiy2010channel}. These bounds can be utilized for the
actual system design in case of the discontinuous transmission with an arbitrary packet length \cite{durisi2016shortpacket}.
Furthermore, ultrareliable communication with short packets has gained an increased attention
recently, where extremely low packet error rate has become one the main requirements and
challenges \cite{durisi2016toward}, \cite{popovski2018wireless}. Various works aim at optimizing the resource allocation and maximizing
the accuracy of channel estimation under the assumed constraints of ultrareliability and ultralow
latency, cf. \cite{she2017radio}, \cite{mousaei2017optimizing}. Although the main focus of the research in this context is clearly on the downlink system design, the problem of ultrareliable data uplink has been addressed as well, cf. \cite{she2018joint}, \cite{singh2018contentionbased}. Furthermore, various scenarios have been investigated, e.g. relaying based
transmissions \cite{hu2018relayingenabled}, \cite{gu2018ultrareliable}, and multiple-input multiple-output (MIMO) systems \cite{vu2017ultrareliable}. The authors
have investigated the performance bounds given by achievable data rates and the optimal system
design for such configurations. Moreover, the feasibility of wireless power transfer (WPT) via
short packets has been studied in \cite{souza2018ultrareliable}. However, all these works pose hard constraints on the
scheduling of transmissions, which render the proposed methods not applicable to RMA.\\
An even more challenging problem is to design a simultaneous wireless information and power
transfer (SWIPT) system, which utilizes randomly scheduled short packet transmissions taking
into account the system requirement of ultrareliable symbol detection. This problem has been addressed for the first time in \cite{kisseleff2019ultrareliable}, where a step-by-step design of the ultrareliable SWIPT system based on power splitting has been shown. Here, the ultrareliability condition has been defined with respect to the worst-case signal quality by providing a lower bound on the instantaneous signal-to-noise ratio (SNR) in each symbol interval. As the most crucial design parameter, the dynamically adjustable power splitting factor (PSF) has been identified, which determines the relative amount of incoming signal power to be fed into the energy harvesting circuit \cite{shi2014joint}. It turns out, that the design problem requires a prediction of the PSF in each consecutive symbol interval, which has been addressed as well using the method proposed in \cite{kisseleff2019ultrareliable} by taking into account the probability of packet arrival and the length of data packets. The corresponding increase of the harvested energy
compared to the naive method without prediction has been demonstrated. This method has been
applied to the target scenario discussed in the following.
\subsection{Scenario}
For the target scenario, we consider a relay-aided uplink of a large communication system,
where a set of nodes is expected to transmit information to a relay node, which may process
the received data by means of redundancy reduction, and then forwards it to the base station
(destination). For this, we assume that any change in the incoming data needs to be detected as soon and as reliably as possible in order to facilitate a quick response of the system. A packet error may lead to an undesired delay or to a possible false alarm. Hence, reliable communication is required in the considered application. Interestingly, due to the assumed RMA, the signal quality for the detection of each packet varies from symbol interval to symbol interval. Hence, it is not possible to minimize the average packet error rate directly. Instead, the detection of each symbol should be made as reliable as possible. This scenario is applicable e.g. in LPWANs, where the nodes are typically assumed
to be far away from the base station. In fact, some of them might be close to each other and would
preferably form a cluster and use a relay-aided (or possibly satellite-aided \cite{kodheli2017integration}) backhaul instead
of disturbing each others' individual transmissions if no relay is employed. Such a scenario is
well known in the field of event-driven wireless sensor networks (cf. \cite{akyildiz2010wireless}), where the cluster
head would only send a short packet to the next cluster or its parent node, if the sensed data is
sufficiently novel and spatially diverse. Similarly, in our scenario, we assume that the amount of
data forwarded by the relay is significantly lower than the total amount of data received by the
relay from the nodes. Correspondingly, the energy consumed by the relay during its transmission
to the destination is also relatively low, such that the relay can even be powered by the received signals\footnote{We assume that the relay device is not too far away from the nodes, such that a reasonable amount of energy can be harvested.} from the surrounding nodes. Note, that we do not consider the power consumption at
the relay despite a potentially extensive processing complexity. This is due to the fact, that the
actual computation can be partially carried out remotely. Interestingly, with increasing number of
nodes and equal probability of transmission, the average amount of harvested energy increases,
since the variance of the received signal increases as well. On the other hand, more and more
adjacent transmissions interfere with each other and reduce the signal quality, such that the
communication becomes unreliable.
\subsection{Contributions}
In this work, we focus on the design of the relay device. We select the most promising
design strategies and provide methods for the optimization of the key system parameter, which
is the dynamically adjustable PSF in our scenario. In this context, a practical method based on
prediction of the best PSF at the relay is proposed. This method maximizes the
harvested energy and guarantees reliable signal acquisition at the same time. Furthermore,
it significantly outperforms the naive baseline scheme.\\
Our contributions comprise:
\begin{itemize}
\item predictor design in presence of a prediction delay, which results from the optimization of the PSF as part of the prediction. In order to cope with the
prediction delay, we propose a novel block-based predictor (BBP), which can be viewed as a generalization of the symbol-based predictor (SBP) proposed in \cite{kisseleff2019ultrareliable};
\item performance analysis under imperfect channel state information (CSI). Here, we take into account a possible deviation of the complex-valued channel gains from the assumed ones
in the predictor design and show that the proposed methods are more robust against the CSI uncertainty than the baseline scheme;
\item complexity analysis of both proposed methods (symbol-based and block-based predictors). In addition, we investigate the influence of the duty cycle on the required number of
multiplications in each symbol interval and explain the possibility of remote computation of the PSF;
\item algorithmic representation of the proposed methods.
\end{itemize}
This paper is organized as follows. The system model with respect to information and energy
transmission as well as reliable signal detection at the relay are discussed in Section II. In Section
III, the problem of maximizing the average harvested power is presented. Also, a practical method
based on state prediction is proposed. Numerical results are shown in Section IV and subsequently
the paper is concluded in Section V.
\section{System Model}
We assume $N$ stationary transmitter devices, e.g. IoT nodes, deployed in a static environment
in close proximity of the relay. Due to the stationary deployment, communication channels
between the nodes and the relay are static, such that a sufficiently good level of synchronization
and availability of the channel state information at the relay can be assumed. Each node (as well
as the relay) is equipped with a single omnidirectional antenna. The relay detects the symbols of transmitted data
packets from all nodes, restructures the data\footnote{This step may include redundancy reduction, data aggregation, compression, decoding and re-encoding, or symbol remapping.}, and forwards it to the destination. The network
structure is depicted in Fig. \ref{fig1}. 
\ifCLASSOPTIONdraftcls
\begin{figure}
\centering
\includegraphics[width=0.6\textwidth]{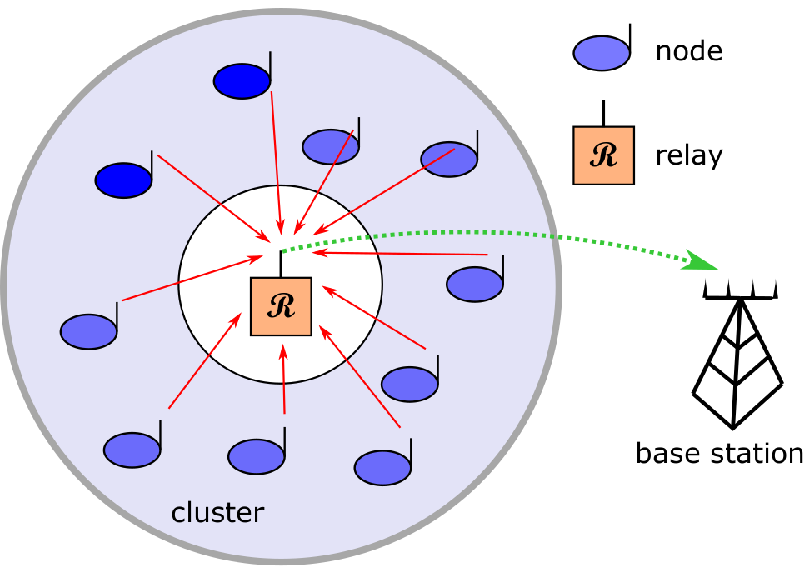}
\caption{Network structure. The nodes are scattered around the relay. The relay uses a separate channel in order to forward the
data to the base station.}
\label{fig1}
\end{figure}
\else
\begin{figure}
\centering
\includegraphics[width=0.42\textwidth]{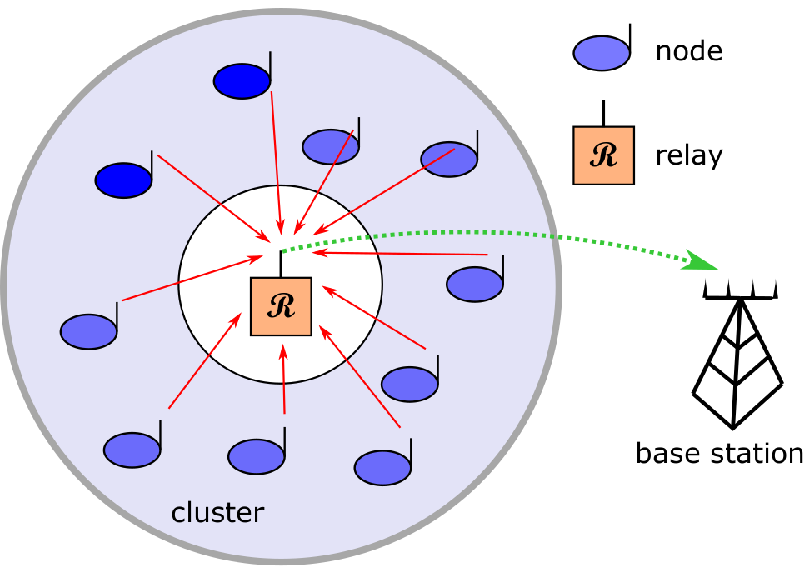}
\caption{Network structure. The nodes are scattered around the relay. The relay uses a separate channel in order to forward the
data to the base station.}
\label{fig1}
\end{figure}
\fi
In addition, the relay may harvest energy from the received
signals. In this work, we focus on the design of the energy harvesting relay for the described
scheme, which guarantees reliable symbol detection. The link between the relay and
the destination remains beyond the scope of this work.\\
For the transmit signal, we assume that each node $n$ decides to transmit a new data packet of length $L_n$ in each symbol interval of length $T$ with probability $p_n$. Note, that a new packet
transmission cannot start during an ongoing packet transmission of the same node. Furthermore,
we assume that the parameters $T$, $p_n$, and $L_n$, $\forall n$ are known to the receiver, e.g. as part of a
standard-compliant system configuration. In particular, $p_n$ is related to the individual duty cycle of node $n$ and can be initialized either by the node or by the relay depending on the priority of the sensing information from the particular node or on the channel state. Furthermore, the packet length $L_n$ can be selected with respect to the performance of the FEC coding and is either considered to be equal to the FEC packet length or to the length of a sub-packet, if methods like Telegram Splitting \cite{6525243} are applied. The knowledge of these parameters can be exploited
in order to improve the system performance as we demonstrate below.\\
For simplicity, the transmit power $P_t$ is equal for all nodes during the packet transmission.
Obviously, in absence of data to be transmitted, i.e. in sleep mode, the power consumption at
the nodes is negligible. Hence, the average consumed power is less than $P_t$
\begin{equation}
\label{eq:cons}
P_{\mathrm{consumed},n} = P_t\frac{p_nL_n}{p_nL_n + (1 - p_n)\cdot 1},\:\forall n,
\end{equation}
since $L_n$ symbol intervals are occupied with probability $p_n$ and one symbol interval is left empty (without actual symbol) with probability $1 - p_n$. This estimate of the average consumed power is based solely on the transmit power during the active mode. A more detailed modeling would include the power consumption during the sleep mode and during the transition from sleep mode into active mode \cite{1532220}. However, since the focus of this work is on the design of the relay (not the transmitters), the simplified power consumption model provided in \eqref{eq:cons} is sufficient for our calculations.\\
The sequence of bits to be transmitted by each node is modulated via coded binary phase-shift
keying\footnote{Such a low modulation rate of only 1 bit/symbol has been assumed in order to account for the use of cheap low-power sensor nodes and in order to increase the reliability of transmission. However, the methods proposed in this paper are applicable to other kinds of modulation including amplitude-shift keying (ASK), higher-order phase-shift keying (PSK) and quadrature amplitude modulation (QAM) as well.} (BPSK), such that a sequence\footnote{Since we consider a discontinuous transmission, we define the data of each packet in the range $0 < m \leq L_n$ and set $c_{n,k}[m] = 0$ otherwise.} $c_{n,k}[m] \in \{-1,+1\}$, $0 < m \leq L_n$ results for each
packet $k$ of node $n$. In addition, a random spacing $\nu_{n,k}$ between packet $k-1$ and $k$ is introduced
according to the underlying probability of transmission $p_n$ for node $n$. Hence, each node $n$
transmits an infinite sequence of randomly shifted data packets
\begin{equation}
a_n[m] =\sum_{k=-\infty}^{\infty}c_{n,k}[m - ((k - 1)L_n + \nu_{n,k})],
\end{equation}
such that 
\begin{equation}
\label{eq:cons_simpl}
P_{\mathrm{consumed},n}=P_t\mathcal{E}_m\{|a_n[m]|^2\}
\end{equation}
holds and $\mathcal{E}_m\{\cdot\}$ denotes the expectation operator with respect to the
received symbols in all symbol intervals $m$ from the underlying random process.
For the signal propagation between the nodes and the relay, we assume frequency-flat quasi-static
block fading with the complex-valued channel coefficient $h_n$, $\forall n$. Also, the channel coefficient
$h_n$ is assumed to be known to the receiver, which is a reasonable assumption for a stationary
deployed network, as mentioned earlier. The received signal is given by
\begin{equation}
y[m] =\sqrt{P_t}\sum_{n=1}^N h_na_n[m] + w[m],
\end{equation}
where $w[m]$ is the additive white Gaussian noise with variance $\sigma^2$. In presence of external interference from other communication systems, the total disturbance variance, which consists of both noise and interference variance, should be employed instead of $\sigma^2$.\footnote{This suggestion is valid in case of continuous signal transmission with a constant variance of disturbance. In presence of bursty interference, the methods proposed in this paper need to be combined with the estimation of the interference state according to \cite{kisseleff2018optimal}.}\\
In the following, we consider the process of packet arrival and its implications for energy
harvesting and symbol detection.
\subsection{Information and energy reception}
There are two major (classical) methods for energy harvesting in SWIPT (cf. \cite{zhang2013mimo}): time splitting (TS) and
power splitting (PS). In the TS approach, the received signal is alternatingly used for information
and energy reception. TS is typically employed in scheduled access based communication
networks, since TS can be viewed as a special case of scheduling of information and energy
transmission. Hence, the use of TS in scheduled access schemes provides a certain level of design
flexibility. In the PS approach, the signal is split by a power splitter, such that one part of the
signal is used for symbol detection and another part of the signal is used for energy harvesting.
For the considered scenario, the TS approach seems to be unfeasible, since some of the nodes
may start their transmission during the energy harvesting phase, such that the respective symbols
of their packets cannot be detected by the receiver. Hence, PS approach is selected.\footnote{In the recent time, there have been attempts to design specific modulation schemes, in particular based on multitone excitation and nonlinear signal amplification, which aim at maximizing the efficiency of SWIPT, cf. \cite{8330063, claessens2019multitone}. However, the applicability of these schemes in the context of multiple access and especially RMA is unknown. Hence, we focus on the classical methods in this work.}
A basic structure of the employed SWIPT module is depicted in Fig. \ref{fig2}. 
\ifCLASSOPTIONdraftcls
\begin{figure}
\centering
\includegraphics[width=0.5\textwidth]{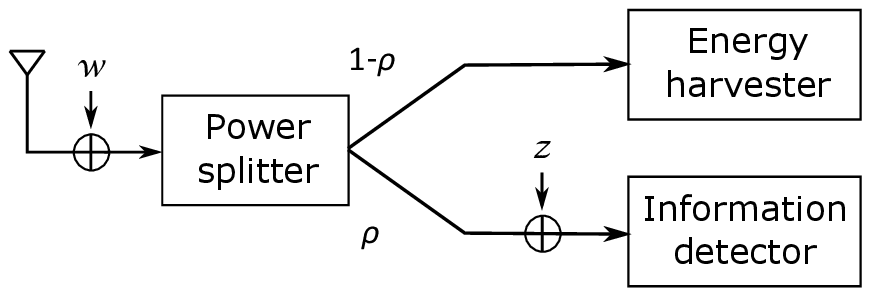}
\caption{Power splitting in SWIPT module.}
\label{fig2}
\end{figure}
\else
\begin{figure}
\centering
\includegraphics[width=0.38\textwidth]{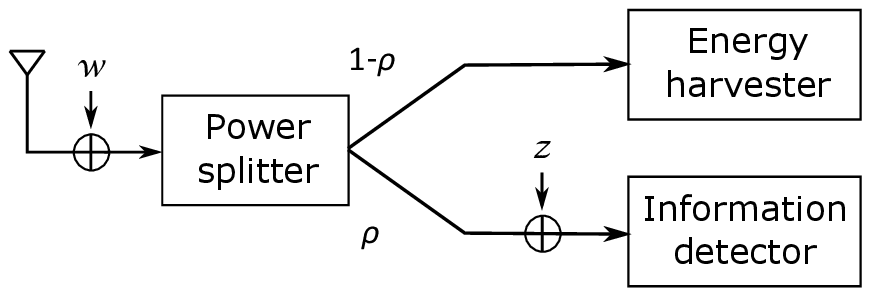}
\caption{Power splitting in SWIPT module.}
\label{fig2}
\end{figure}
\fi
Typically, the splitting
of the signal power results in a decrease of the signal quality given by the signal-to-noise
ratio (SNR) at the input of the symbol detector. We model this degradation by adding a white
Gaussian noise signal $z(t)$ with the variance $\delta^2$ to the received signal, cf. \cite{shi2014joint}. 
Due to the discontinuous transmission via short packets, the harvested energy fluctuates depending
on the presence or absence of the signals from the individual nodes. The mean harvested power is given by\footnote{For the derivation of \eqref{eq:harv_lin}, we assume uncorrelated symbols from all $N$ nodes, which is partially motivated by the RMA.}
\ifCLASSOPTIONdraftcls
\begin{eqnarray}
\label{eq:harv_lin}
P_{\mathrm{harv}}\equalalign\mathcal{E}_m\{(1-\rho)\eta |y[m]|^2\}=(1-\rho)\eta\mathcal{E}_m\left\{\left|\sum_{n=1}^N \sqrt{P_t}h_na_n[m] + w[m]\right|^2\right\}\notag\\
\equalalign(1-\rho)\eta\sum_{n=1}^N \left(|h_n|^2P_t\mathcal{E}_m\{|a_n[m]|^2\} + \sigma^2\right)=(1-\rho)\eta \left(\sum_{n=1}^N\frac{P_t|h_n|^2 p_nL_n}{p_nL_n + (1 - p_n)}+\sigma^2\right),
\end{eqnarray}
\else
\begin{eqnarray}
\label{eq:harv_lin}
P_{\mathrm{harv}}\equalalign\mathcal{E}_m\{(1-\rho)\eta |y[m]|^2\}\notag\\
\equalalign(1-\rho)\eta\mathcal{E}_m\left\{\left|\sum_{n=1}^N \sqrt{P_t}h_na_n[m] + w[m]\right|^2\right\}\notag\\
\equalalign(1-\rho)\eta\sum_{n=1}^N \left(|h_n|^2P_t\mathcal{E}_m\{|a_n[m]|^2\} + \sigma^2\right)\notag\\
\equalalign(1-\rho)\eta \left(\sum_{n=1}^N\frac{P_t|h_n|^2 p_nL_n}{p_nL_n + (1 - p_n)}+\sigma^2\right),
\end{eqnarray}
\fi
where $\rho$ stands for the PSF. Furthermore, we assume $10^{-2} \leq \rho \leq 1$, where $10^{-2}$ is selected as the lower bound on $\rho$, since at least a small
amount of power is needed for the information detection\footnote{This lower bound seems reasonable, since an even lower value $\rho=10^{-3}$ would increase the harvested power very insignificantly, i.e. by less than $1\%$.}. In addition, $\eta$ is the efficiency of conversion of the received signal into electrical energy. Note, that in \eqref{eq:harv_lin}
we apply a linear energy harvesting model with a constant $\eta$, since we assume that the operating point of the energy harvester is in its linear region. The impact of the non-linear behavior of
energy harvesting circuits can be modeled using equations provided in \cite{boshkovska2015practical} or \cite{clerckx2018beneficial}. With the non-linear model from \cite{boshkovska2015practical}, the mean harvested power is given by 
\begin{equation}
\label{eq:harv_nonlin}
P_{\mathrm{harv}}=\mathcal{E}_m\left\{\frac{\frac{\varphi(1+\mathrm{exp}(\psi\phi))}{1+\mathrm{exp}(-\psi((1-\rho)|y[m]|^2-\phi))}-\varphi}{\mathrm{exp}(\psi\phi)T}\right\},  
\end{equation}
where $\varphi$, $\psi$, and $\phi$ are the parameters of the energy harvesting circuit, which can be found via curve fitting \cite{boshkovska2015practical}. As we show in the next section, for the considered application and objective of this work, the system design is independent from the energy harvesting model. Hence, for the clarity of exposition, we assume a linear harvesting model in \eqref{eq:harv_lin} instead of \eqref{eq:harv_nonlin} in the following.\\ 
For the information detection, the average SNR after the power splitting is given by
\ifCLASSOPTIONdraftcls
\begin{equation}
\label{SNR_av}
\mathrm{SNR}_\mathrm{average}=\frac{\rho\mathcal{E}_m\left\{\left|\sum_{n=1}^N \sqrt{P_t}h_na_n[m]\right|^2\right\}}{\mathcal{E}_m\left\{\left|\sqrt{\rho}w[m]+z[m]\right|^2\right\}}=\frac{\rho P_t\sum_{n=1}^N|h_n|^2\frac{p_nL_n}{p_nL_n + (1 - p_n)}}{\rho\sigma^2+\delta^2}.
\end{equation}
\else
\begin{eqnarray}
\label{SNR_av}
\mathrm{SNR}_\mathrm{average}\equalalign\frac{\rho\mathcal{E}_m\left\{\left|\sum_{n=1}^N \sqrt{P_t}h_na_n[m]\right|^2\right\}}{\mathcal{E}_m\left\{\left|\sqrt{\rho}w[m]+z[m]\right|^2\right\}}\notag\\
\equalalign\frac{\rho P_t\sum_{n=1}^N|h_n|^2\frac{p_nL_n}{p_nL_n + (1 - p_n)}}{\rho\sigma^2+\delta^2}.
\end{eqnarray}
\fi
However, due to the discontinuous packet transmissions, the instantaneous SNR 
\begin{equation}
\mathrm{SNR}_\mathrm{instant}[m]=\frac{\rho P_t\left|\sum_{n=1}^{N}h_na_n[m]\right|^2}{\rho\sigma^2+\delta^2}.
\end{equation}
in each symbol interval $m$ might be either lower or higher than the average SNR given in \eqref{SNR_av} depending on the number of active nodes. In fact, if $\mathrm{SNR}_\mathrm{instant}[m]$ is lower than $\mathrm{SNR}_\mathrm{average}$ due to the collisions of multiple packets from adjacent transmissions, symbol errors may occur, which may significantly degrade the system performance. On the other hand, if $\mathrm{SNR}_\mathrm{average}$ is lower than $\mathrm{SNR}_\mathrm{instant}[m]$, then the PSF $\rho$ is not properly chosen, since too much energy is put into the information detection and correspondingly less energy is harvested. If we assume a constant PSF, which guarantees a highly reliable symbol detection in all symbol intervals, the harvested energy will always be extremely low. In many cases, the energy harvesting may even be unfeasible. However, it is not possible to obtain a better solution with larger average harvested power using a constant PSF without violating the imposed requirements of signal quality.\\
In this work, we assume (similar to \cite{liu2013chua}) that the PSF can be dynamically adjusted in order to account for the time-varying receive power and the interference from adjacent transmissions. Correspondingly, we denote $\rho[m]$ as the PSF that is used in the $m$th symbol interval. Note, that the PSF has to be known before symbol detection, since information detection and further processing is done after the splitting, see Fig. 2. Hence, $\rho[m]$ needs to be predicted before the respective symbol interval. Assuming that $\rho[m]$ can be predicted during the reception of the previous symbol, such that it can be updated before the reception of the next symbol, a symbol-based predictor has been proposed in \cite{kisseleff2019ultrareliable}. This prediction is based on the estimation of transmission probability for each node in the next symbol interval using the observations of previous symbols. Accordingly, the optimal $\rho[m]$ is determined under the ultrareliability constraint using the combinations of signals from the nodes with a non-vanishing probability of transmission.\\
Given the large number of nodes, which can influence the prediction, the calculation of the
optimal PSF may require a substantial computational effort. For this calculation,
more time than just one symbol interval may be needed. Hence, we can assume that an update
of $\rho[m]$ requires $D$ symbol intervals. Since the next update is only possible after the
processing of subsequent $D$ symbols\footnote{Unfortunately, it is not possible to update $\rho[m]$ in every symbol interval, since we assume that a new calculation can only start after $D$ symbols. Hence, a "pipelining" based processing via e.g. a shift register would lead to a performance degradation, since the processed symbols will become more and more outdated with each new calculation.}, the splitting factor should remain unchanged for the next
$D$ symbols. Correspondingly, the prediction is done for a block of $D$ symbols, such that we
obtain a BBP. Note, that for the prediction, we exploit a long-term observation,
which is significantly longer than $D$. However, it is not possible to apply a sequence estimation
here, since parts of the packet may be missing and this would lead to an incorrect prediction of
the PSF.\\
One of the problems of the BBP is due to the fact that no additional symbols
can be taken into account during the calculation of $\rho[m]$, see Fig. \ref{fig3}. 
\ifCLASSOPTIONdraftcls
\begin{figure}
\centering
\includegraphics[width=0.7\textwidth]{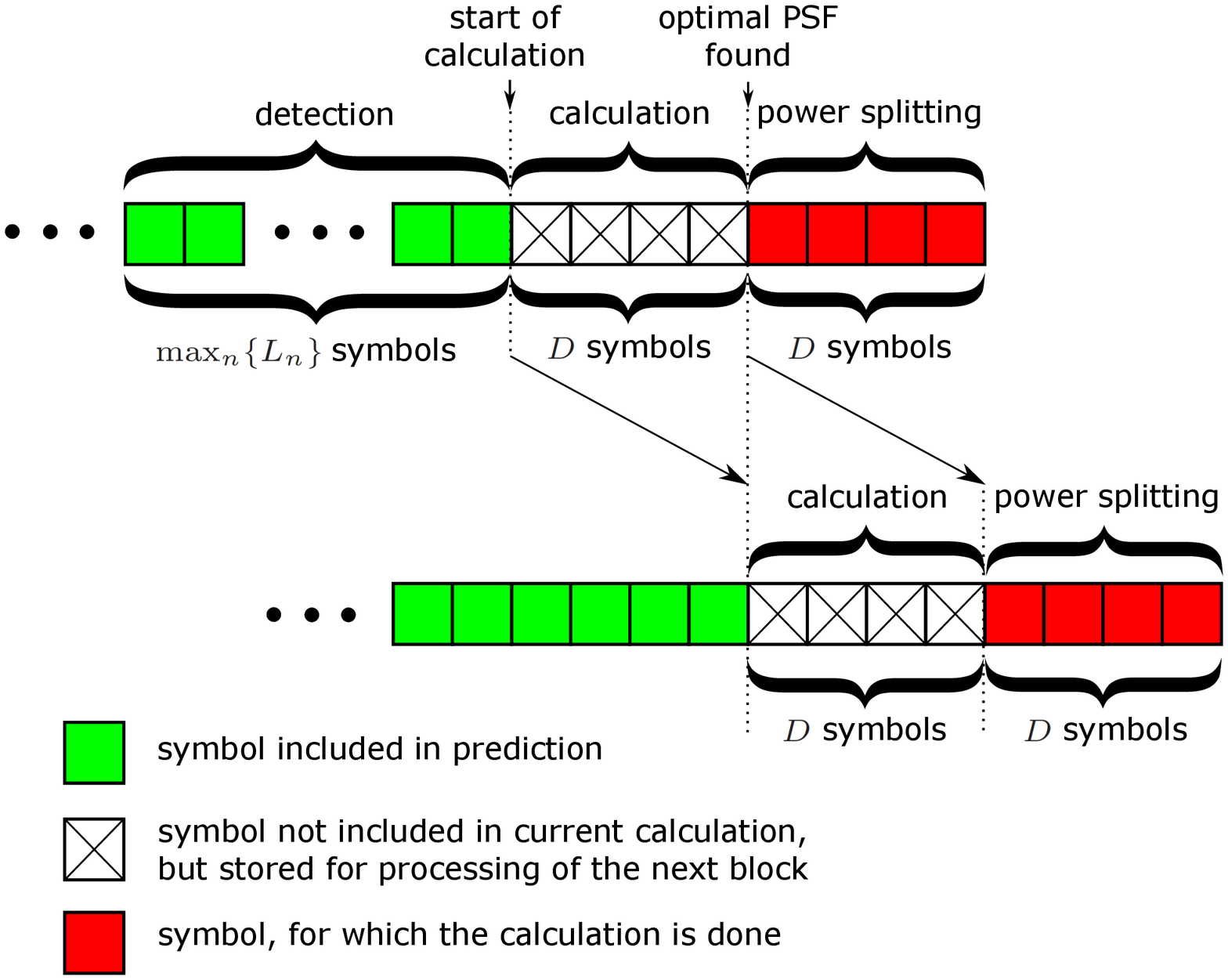}
\caption{Prediction of two consecutive blocks. Calculation of $\rho[m]$ requires a prediction of $2D$ symbols.}
\label{fig3}
\end{figure}
\else
\begin{figure}
\includegraphics[width=0.49\textwidth]{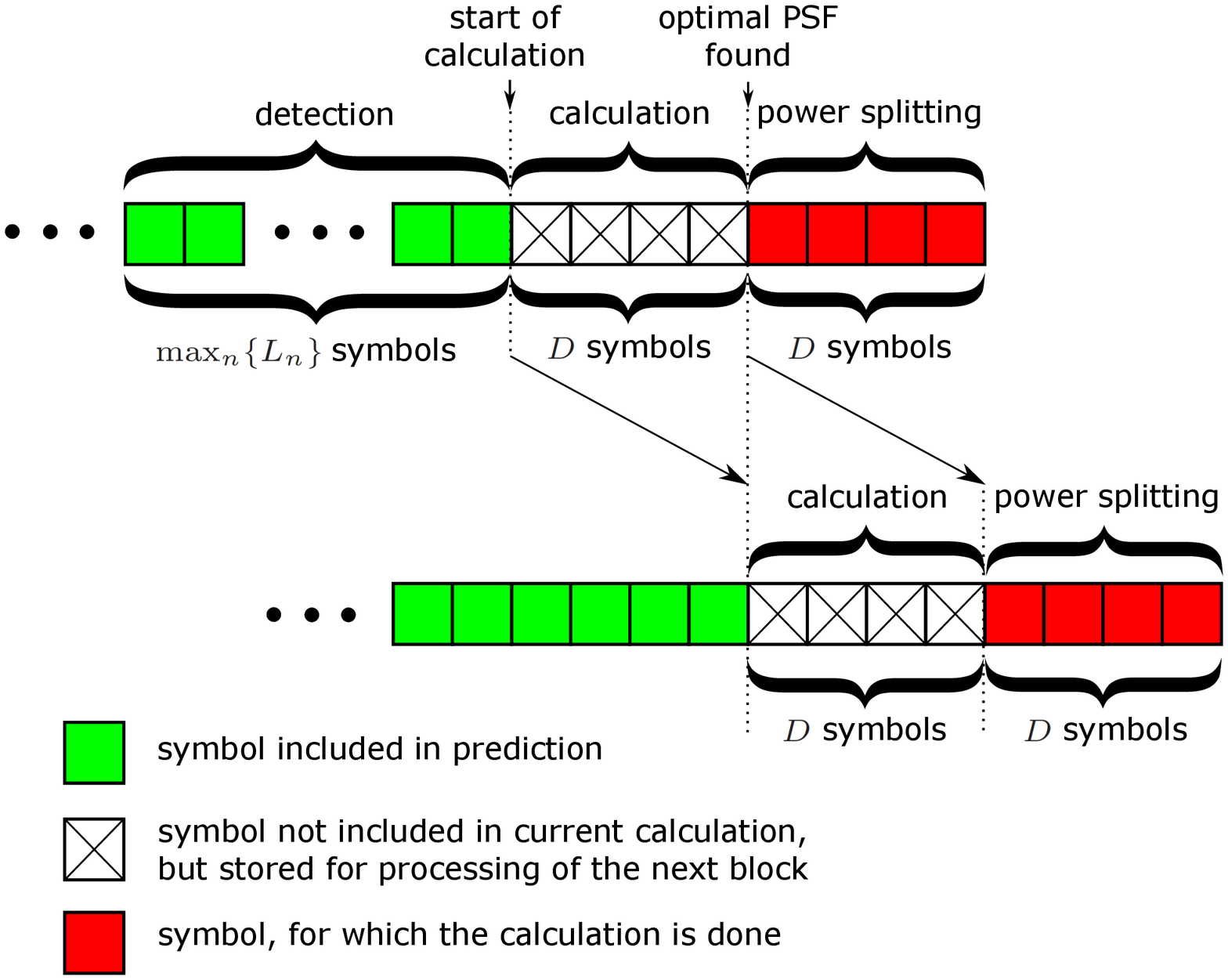}
\caption{Prediction of two consecutive blocks. Calculation of $\rho[m]$ requires a prediction of $2D$ symbols.}
\label{fig3}
\end{figure}
\fi
Apparently, the first symbol
from the symbol block, for which the prediction is done, lies $D$ symbols ahead of the last
symbol that is taken into account in the calculation. The last symbol, for which the calculation is done, lies $2D$ symbols ahead. Correspondingly, in order to predict the states for all $D$ symbols
of the target block, we need to predict $2D$ symbols, which leads to a significant performance
degradation compared to the SBP. Furthermore, this calculation is much more
computationally complex, such that a trade-off between complexity and accuracy of prediction
results. An insight into this trade-off is provided by the complexity analysis in Section \ref{sec_complexity}.
\subsection{Reliable detection}
There are different methods of retrieving the transmitted data of all packets from the received
signal $y[m]$. Among others, successive interference cancellation (SIC) and joint detection (JD) are
the most popular ones. These methods (especially SIC) are widely used in the context of NOMA
in order to separate overlapping data streams [4]. While JD is optimal for a symbolwise signal
detection, SIC is beneficial in sequence detection, since the dependencies among the individual
symbols (attributed e.g. to the channel memory or coding) can be exploited in order to increase
the detection performance. In the considered scenario, the symbol detection cannot wait for the
whole data packet to be received, since the PSF needs to be adjusted before
the reception of each symbol or a block of symbols, as mentioned earlier. Correspondingly, it
is difficult to exploit the dependencies among the symbols. In addition, SIC performs well only in case of sufficiently separable symbol streams, e.g. in terms of received signal power. In our scenario, there may be no dominant signal power or it may correspond to a very short part of the transmitted sequence, e.g. in the beginning of a transmission, such that not enough information is collected for the accurate interference cancellation. Hence, we select JD for symbol detection.\\
In the JD approach, a new constellation of signal points is created, which results from combining
the signal points of all involved transmissions weighted with the respective channel coefficients.
As an example, assume that two nodes transmit individually or simultaneously sequences of
BPSK symbols, which pass through the individual channels $h_1$ and $h_2$.\footnote{For the clarity of exposition, we set the transmit power to 1.} Consider the $m$th symbol
interval. If only the first or the second node transmits, the constellation points are $\{-h_1,+h_1\}$
or $\{-h_2,+h_2\}$, respectively. If both of them transmit, there are four points in a joint signal
constellation: $\{-h_1 - h_2,+h_1 -h_2,-h_1 + h_2,+h_1 + h_2\}$. Obviously, a symbol error can occur,
if the noise signal is stronger than half of the minimum Euclidean distance $d[m]$ between any
two constellation points of the new constellation. Note, that $d[m]$ depends on $\rho[m]$ in terms
of $d[m] = d_0[m]\sqrt{\rho[m]}$, where $d_0[m]$ is the normalized minimum distance between any two
constellation points. When only two nodes are active in a particular symbol interval, a symbol error in JD may
potentially result in a symbol error in each of the respective packets of both nodes. With
increasing number of nodes, the impact of a symbol error in JD becomes very high and may
render the packet detection impossible. This issue is especially crucial for the beginning of
a new packet transmission, which can be missed in case of wrong detection. Such a missed
detection may result in a shift of the data within the packet, such that the whole packet would
be damaged. Moreover, an erroneous packet detection may lead to error propagation from packet
to packet and damage the reception of all subsequent packets. In order to avoid the packet loss and the error propagation, we design the signal detector
according to a conservative upper bound on the overall symbol error rate. For this, we consider
the signal quality $\mathrm{SNR}_\mathrm{mod}[m]$ observed at the information detector with respect to the most vulnerable constellation points. This strategy is sometimes used in case of a non-trivial multiuser
detection \cite{verdu1998cambridge}. Hence, we determine the Euclidean distance between any two points of the joint constellation and select the minimum distance among all point pairs. We define the modified
SNR as
\begin{equation}
\label{eq:6}
\mathrm{SNR}_\mathrm{mod}[m]=\frac{(0.5d[m])^2}{\delta^2+\rho[m]\sigma^2}=\frac{(0.5d_0[m])^2\rho[m]}{\delta^2+\rho[m]\sigma^2},
\end{equation}
where $\mathrm{SNR}_\mathrm{mod}[m]$ depends on the constellation and correspondingly on the presence of packets from the active nodes in the $m$th symbol interval.\\
In order to guarantee a sufficiently reliable symbol detection, we assume $\mathrm{SNR}_\mathrm{mod}[m] \geq 13$dB $\triangleq 20$. Through this, the upper bound of the symbol error probability becomes very small
according to \cite{goldsmith2005cambridge} and the communication is reliable. By inverting \eqref{eq:6} and using the definition of $\rho$, we obtain
\begin{equation}
\label{eq:rho}
\max\Big\{\frac{20\delta^2}{(0.5d_0[m])^2-20\sigma^2}, 10^{-2}\Big\}\leq \rho[m]\leq 1.
\end{equation}
Moreover, $\rho[m]$ remains constant for $D$ symbol intervals in case of a BBP,
as mentioned earlier. In order to model this behavior, we introduce a starting index $q(m) =
m-\mathrm{mod}_D(m)$ for the respective updates in the interval $q(m) \leq m < q(m)+D$, where $\mathrm{mod}_i(j)$
denotes the modulo operation with basis $i$ applied to $j$. Hence, $\rho[m] = \rho[q(m)]$, holds. This
constraint together with (7) will be used for the design of the power splitter in the next section.
\section{Power splitting optimization}
In this section, we address the choice of the optimal PSF $\rho[m]$. We start with
the problem formulation for SBP and BBP. Then we analyze the
performance bounds given by a simple baseline scheme (lower bound) and a genie-aided optimization
(upper bound). After that, practical methods are proposed, which exploit the available
knowledge of signal statistics. Furthermore, the implications for the prediction of the power
splitting factor related to imperfect CSI and the computation complexity are discussed.
\subsection{Prediction problem}
In this work, we would like to explore the potential of SWIPT for reliable unscheduled
short packet transmissions from multiple nodes. In order to account for the varying number of
nodes and the corresponding joint symbol constellation, the PSF needs to be
continuously adapted. Hence, the goal is to find a good sequence $\rho[m]$, $\forall m$. In this paper, we
focus on the average harvested power $P_{\mathrm{harv}}(\rho[m])$ as a performance metric, which is obtained
by taking into account the time-varying PSF in \eqref{eq:harv_lin}. Here, $P_{\mathrm{harv}}(\rho[m])$ indicates
that the choice of the sequence $\rho[m]$ heavily affects the average harvested power. For the SBP, we formulate the following optimization problem using \eqref{eq:rho}:
\ifCLASSOPTIONdraftcls
\begin{eqnarray}
\label{eq:prob1}
\emptyalign\hspace*{-12mm}\max_{\rho[m]}\:\mathcal{E}_m\{(1-\rho[m])\eta |y[m]|^2\},\\
\mbox{s.t.:} \emptyalign C1)\: \max\Big\{\frac{20\delta^2}{(0.5d_0[m])^2-20\sigma^2}, 10^{-2}\Big\}\leq \rho[m]\leq 1,\notag\\
\emptyalign C2)\: d_0[m]\: \mathrm{unknown\: before\: symbol\: interval}\: m+1.\notag
\end{eqnarray}
\else
\begin{eqnarray}
\label{eq:prob1}
\emptyalign\hspace*{-11mm}\max_{\rho[m]}\:\mathcal{E}_m\{(1-\rho[m])\eta |y[m]|^2\},\\
\hspace*{1mm}\mbox{s.t.:} \emptyalign \hspace*{-4mm}C1a) \max\Big\{\hspace*{-0.5mm}\frac{20\delta^2}{(0.5d_0[m])^2-20\sigma^2}, \hspace*{-0.5mm}10^{-2}\Big\}\hspace*{-0.5mm}\leq\hspace*{-0.5mm}\rho[m],\notag\\
\emptyalign \hspace*{-4mm}C1b)\: \rho[m]\leq 1,\notag\\
\emptyalign \hspace*{-4mm}C2)\: d_0[m]\: \mathrm{unknown\: before\: symbol\: interval}\notag\\
\emptyalign \hspace*{4mm}m+1.\notag
\end{eqnarray}
\fi
Obviously, this problem cannot be solved analytically, since the solution to \eqref{eq:prob1} involves a
prediction of $d_0[m]$. Similarly, for the BBP, we formulate the problem
\ifCLASSOPTIONdraftcls
\begin{eqnarray}
\emptyalign\hspace*{-12mm}\max_{\rho[m]}\:\mathcal{E}_m\{(1-\rho[m])\eta |y[m]|^2\},\\
\mbox{s.t.:} \emptyalign C1)\: \max\Big\{\frac{20\delta^2}{(0.5d_0[m])^2-20\sigma^2}, 10^{-2}\Big\}\leq \rho[m]\leq 1,\notag\\
\emptyalign C2)\: \rho[m]=\rho[q(m)],\notag\\
\emptyalign C3)\: q(m) =m-\mathrm{mod}_D(m),\notag\\
\emptyalign C4)\: d_0[m]\: \mathrm{unknown\: before\: symbol\: interval}\: m+1.\notag
\end{eqnarray}
\else
\begin{eqnarray}
\emptyalign\hspace*{-12mm}\max_{\rho[m]}\:\mathcal{E}_m\{(1-\rho[m])\eta |y[m]|^2\},\\
\hspace*{1mm}\mbox{s.t.:} \emptyalign \hspace*{-4mm}C1a) \max\Big\{\hspace*{-0.5mm}\frac{20\delta^2}{(0.5d_0[m])^2-20\sigma^2}, \hspace*{-0.5mm}10^{-2}\Big\}\hspace*{-0.5mm}\leq\hspace*{-0.5mm}\rho[m],\notag\\
\emptyalign \hspace*{-4mm}C1b)\: \rho[m]\leq 1,\notag\\
\emptyalign \hspace*{-4mm}C2)\: \rho[m]=\rho[q(m)],\notag\\
\emptyalign \hspace*{-4mm}C3)\: q(m) =m-\mathrm{mod}_D(m),\notag\\
\emptyalign \hspace*{-4mm}C4)\: d_0[m]\: \mathrm{unknown\: before\: symbol\: interval}\notag\\
\emptyalign \hspace*{4mm}m+1.\notag
\end{eqnarray}
\fi
Note, that these optimization problems are formulated with respect to the employed linear energy harvesting model. However, due to the monotonic decrease of the harvested power with respect to $\rho[m]$ according to \eqref{eq:harv_nonlin}, the solution of the respective optimization problems assuming a non-linear harvesting model would be the same as with the linear model in each symbol interval. In order to tackle these problems, we first consider the feasible performance bounds and then
describe our proposed solutions.
\subsection{Lower and upper bounds}
For the lower bound of the harvested power, we assume that no prediction of $d_0[m]$ is
applied. Correspondingly, $\rho[m]$ is constant and needs to be selected only once. Hence, it may
not be possible to account for the different combinations of packets from various nodes, such
that instead all possible combinations of symbols need to be taken into account in a globally joint symbol constellation. For example, assuming again two nodes with individual constellation
points $\{-h_1,+h_1\}$ and $\{-h_2,+h_2\}$, the resulting globally joint constellation would comprise
the following points: $\{0,-h_1,+h_1,-h_2,+h_2,-h_1-h_2,+h_1-h_2,-h_1+h_2,+h_1+h_2\}$. Since the
maximum number of constellation points is considered in this scheme while the average received
energy is equal for all detection schemes, the minimum distance between the constellation points
is expected to be minimal, such that this scheme provides the lower bound for both $d_0[m]$
and $P_{\mathrm{harv}}(\rho[m])$. We denote this scheme as our baseline scheme. Interestingly, one may try to
combine the baseline scheme with a suitable forward error correction (FEC) coding, since the
PSF does not need to be updated after each symbol interval. The resulting
coding gain, which pertains to the selected FEC method, can be used in order to reduce the
symbol error rate while keeping the relative distance between constellation points somewhat
smaller than without FEC. Through this, the harvested power can be increased. However, no
method of sequence estimation for multiple adjacent transmissions in an RMA configuration
is known to date. Correspondingly, a symbol-by-symbol detection is preferred, which can be
optimally done using JD, as mentioned earlier. Hence, no FEC can be exploited in this case and
the described baseline scheme is a valid benchmark for the performance evaluation.
For the upper bound of the harvested power, we consider an ideal (genie-aided) prediction
of $d_0[m]$. For this, we assume that the receiver knows exactly which nodes transmit in each
symbol interval $m$. Although the actual transmitted symbols are still unknown to the receiver,
this information helps to eliminate most of the constellation points, which pertain to the invalid
combinations of packets. Due to the perfect prediction, this scheme provides a theoretical upper
bound for the system performance. However, this method is impractical, since the knowledge
about the transmissions, which are about to start, is not available in RMA.\\

In this work, we do not address the typical trade-off between information and power transfer,
which is described by rate-energy region (RER), cf. [23]. RER results from the variation of
the PSF, such that the signal quality of the data stream also varies between
very low and very high SNR values. Through this, the joint performance bound in terms of
maximum data rate and harvested energy is provided. However, in order to guarantee reliable
communication, the signal quality needs to be permanently very high, which renders the RER
analysis not feasible. Correspondingly, we focus on the harvested power in this work.
\subsection{Symbol-based predictor (SBP)}
At first, we consider the prediction of $d_0[m]$ using a SBP, i.e. if the
prediction and the update of $\rho[m]$ is possible within one symbol interval. Hence, the predictor
is able to follow all the changes of the time-variant signal quality.
In order to maximize the distance between the constellation points, we try to reduce the number
of points by exploiting the available knowledge on the signal characteristics, in particular the
packet length and the probability of transmission by each node. The idea is to model the useful
part of the received signal as a Markov process. Then, using the current state of the process, the
next state can be predicted. The prediction algorithm is described in Algs. 1 and 2 in Appendix.
We define the state of the Markov process as a vector $\textbf{s}[m]$ of length $N$. The $n$th element of
$\textbf{s}[m]$ is denoted as $s_n[m]$. We set $s_n[m]$ to '1', if $n$th node is currently transmitting, and to '0' otherwise. With this definition, it is possible to obtain the transition probabilities $\mathrm{Pr}(\textbf{s}[m]\:|\:\textbf{s}[m-1])$ from state $\textbf{s}[m-1]$ to state $\textbf{s}[m]$. Each transition probability depends on the probabilities
$\mathrm{Pr}(s_n[m]\: |\: s_n[m - 1]),\: \forall n$. In order to calculate these probabilities, we distinguish between four cases:
\begin{enumerate}
\item $s_n[m-1] = 0$ and $s_n[m] = 0$. Apparently, $\mathrm{Pr}(s_n[m]\: |\: s_n[m-1]) = 1-p_n$ holds, since the $n$th node has decided to not start a new transmission;
\item $s_n[m - 1] = 0$ and $s_n[m] = 1$. In this case, we set $\mathrm{Pr}(s_n[m]\: |\: s_n[m - 1]) = p_n$, since the $n$th node has decided to start a new packet transmission;
\item $s_n[m-1] = 1$ and $s_n[m] = 0$. This case can only occur at the end of the packet transmission.
Hence, a sequence of elements $[s_n[m-L_n], s_n[m- L_n + 1], \ldots, s_n[m-2]]$ is considered
in order to check if the packet transmission has ended. We distinguish between two (sub-)cases:
\begin{enumerate}
\item $s_n[m-l] = 1,\: \forall 2 \leq l \leq L_n$. A packet transmission is finished and the transmitter can
again decide to transmit or not to transmit a new packet. In this case, $\mathrm{Pr}(s_n[m]\:|\: s_n[m-
1]) = 1 - p_n$, since a new transmission has not started;
\item $s_n[m-l] = 0,\: l < L_n$. Since the transmission of the packet cannot be stopped before
the packet end, we set $\mathrm{Pr}(s_n[m]\: |\: s_n[m - 1]) = 0$.
\end{enumerate}
\item $s_n[m - 1] = 1$ and $s_n[m] = 1$. This situation can occur in two (sub-) cases:
\begin{enumerate}
\item during the packet transmission, i.e. $s_n[m - l] = 0,\: l < L_n$. Then, $\mathrm{Pr}(s_n[m]\: |\: s_n[m -1]) = 1$, since the transmission would not stop in the subsequent symbol interval;
\item if a new packet transmission starts directly after the end of the previous packet, i.e.
$s_n[m - l] = 1,\: \forall 2 \leq l \leq L_n$. Then, $\mathrm{Pr}(s_n[m]\: |\: s_n[m - 1]) = p_n$ holds.
\end{enumerate}
\end{enumerate}
The overall transition probability $\mathrm{Pr}(\textbf{s}[m]\: |\: \textbf{s}[m - 1])$ is obtained by multiplying the individual transition probabilities $\mathrm{Pr}(s_n[m]\: |\: s_n[m - 1])$:
\begin{equation}
\mathrm{Pr}(\textbf{s}[m]\: |\: \textbf{s}[m - 1])=\prod_{n=1}^N\mathrm{Pr}(s_n[m]\: |\: s_n[m - 1]),
\end{equation}
since the packet transmissions from different nodes follow independent processes of packet
generation. Through this, we obtain the transition probabilities between the currently observed\footnote{Due to a highly reliable detection with $\mathrm{SNR}_\mathrm{mod}[m]\geq 13$ dB, the state $\textbf{s}[m - 1]$ can be reliably identified.} state $\textbf{s}[m - 1]$ and any other state $\textbf{s}[m]$. Since state $\textbf{s}[m - 1]$ is known, $\mathrm{Pr}(\textbf{s}[m]\: |\: \textbf{s}[m-1])$ is
equal to the probability $\mathrm{Pr}(\textbf{s}[m])$ of occurrence of the respective state $\textbf{s}[m]$ in symbol interval $m$. Moreover, based on the calculated probabilities, we select the most likely states, e.g. according
to $\mathrm{Pr}(\textbf{s}[m]) \geq 10^{-8}$, which need to be taken into account in the joint symbol constellation. These short-listed states are considered in more detail.\\
Each state corresponds to a set of constellation points, which result from the overlap of the
individual symbols of the active nodes according to the respective vector $\textbf{s}[m]$. As an example
with two nodes, $\textbf{s}[m] = [1, 0]$ indicates that only the first node is active and the constellation
comprises the points $\{-h_1,+h_1\}$, whereas $\textbf{s}[m] = [1, 1]$ represents a simultaneous transmission from both nodes and the constellation comprises the points $\{-h_1-h_2,+h_1-h_2,-h_1+h_2,+h_1+
h_2\}$. In case of all-zero vector $\textbf{s}[m] = [0, 0]$, the constellation contains only one point, which is $\{0\}$. Correspondingly, for each state $\textbf{s}[m]$ that has been short-listed according to its probability of occurrence, the constellation points are collected. All these points are likely to be observed in the $m$th symbol interval and are therefore part of a large joint symbol constellation. Obviously,
the number of points in this constellation is smaller than the number of points in the constellation
of the baseline scheme. Hence, a gain in terms of the minimum distance and harvested energy
compared to the baseline scheme can be expected.\\
Furthermore, the distances between any two symbols of the constellation are calculated and the
minimum distance $d_0[m]$ is obtained. Then, the lowest $\rho[m]$ in the range $\Big[\max\Big\{\frac{20\delta^2}{(0.5d_0[m])^2-20\sigma^2}, 10^{-2}\Big\}, 1\Big]$ is selected. Through this, $P_{\mathrm{harv}}(\rho[m])$ in (8) is maximized.\\
As the focus of this paper is on the maximization of the harvested power, we assume that no prediction errors occur due to very high $\mathrm{SNR}_{\mathrm{mod}}$. However, a prediction error can impact multiple subsequent predictions, since the predicted constellation in the each symbol interval depends on the previous predictions. Correspondingly, we suggest to reset the prediction after a certain period of time in order to avoid error propagation and apply a PSF according to the baseline scheme. Of course, in such symbol intervals, the harvested energy would become very low according to the baseline performance. However, due to the reliability condition, the prediction errors are extremely rare, such that the reset procedure can be made rare as well. Correspondingly, the performance degradation with respect to the harvested energy is negligible in this case.
\subsection{Block-based predictor (BBP)}
For the BBP, we need to predict the states for $2D$ symbol intervals using the
previously observed symbol sequence. This prediction can be done iteratively according to Alg.
3 in Appendix. The idea is to consecutively predict the states for each symbol interval $m$ using
Alg. 2 based on each likely state, which results from the prediction for the previous symbol
interval $m - 1$. In contrast to SBP, BBP cannot rely on a
known (already decided) state $\textbf{s}[m - 1]$ as described previously, such that
\begin{equation}
\mathrm{Pr}(\textbf{s}[m]) = \mathrm{Pr}(\textbf{s}[m]\:|\:\textbf{s}[m-1])\mathrm{Pr}(\textbf{s}[m-1]))
\end{equation}
holds and only the states $\textbf{s}[m]$ with $\mathrm{Pr}(\textbf{s}[m]) \geq 10^{-8}$ according to (11) are considered. Hence, the number of states taken into account in the calculation of the minimum distance $d_0[m]$ remains relatively low compared to the baseline scheme despite an increased uncertainty due to a long term
($2D$ symbols) prediction.\\
The resulting likely constellations from all relevant states are combined in order to form a
joint constellation, which is stored in $\textbf{u}_j$ for each symbol interval $j$ of the block. Then, the
minimum distance $d_0[j]$ between the constellation points is calculated for each symbol interval.
The minimum distance $d_0[q(m)] = \min_j d_0[j]$ among all calculated distances is then used for the
calculation of the maximum splitting factor $\rho[q(m)]$. Obviously, this splitting factor is sufficiently
large in order to reliably distinguish the symbols in each symbol interval in the given range.
\subsection{Imperfect CSI}
In the previous sections, we assumed that channel estimation has been thoroughly carried out as part of the receiver synchronization, which precedes the start of data transmission. Typically, a sufficient level of synchronization can be achieved for stationary IoT nodes, as mentioned earlier, such that highly accurate CSI is realistic. However, some of the nodes may not have perfect synchronization due to hardware imperfections or limited channel estimation capabilities. In fact, each new node can introduce additional uncertainty into the signal detection, which may substantially impact the reliability of communication and the prediction of the optimal PSF.\\ 
If a pilot-based estimator is employed, which minimizes the mean-squared error (MSE), the possible imperfections of the CSI can be directly deduced from the well-known performance of this estimator \cite{meyr1997digital}. In order to incorporate the uncertainty related to imperfect CSI, we model the transmission channels as Gaussian-distributed random variables $\hat{h}_n = \bar{h}_n +\tilde{h}_n$ for each node $n$ with respective mean values $\bar{h}_n = \mathcal{E}_m\{\hat{h}_n\}$ and variances $\mathcal{E}_m\{|\tilde{h}_n|^2\} = \vartheta^2_n$, where $\vartheta^2_n$ is typically a small percentage of $|\bar{h}_n|^2$. Since the performance of the traditional channel estimators is known, the variance $\vartheta^2_n$ can be determined based on the length of the employed training sequence for a given estimation approach \cite{mengali1997synchronization}. Hence, we assume that $\vartheta^2_n.\:\forall n$ is known to the receiver.\\
Moreover, we assume that only a small number $N_{ah}$ of ad hoc nodes have a considerable channel variance, i.e. $\vartheta^2_n > 0,\: \forall 1 \leq n \leq N_{ah}$. All other $N - N_{ah}$ nodes are assumed to have
communication channels, which are perfectly known to the receiver, i.e. $\vartheta^2_n = 0,\: \forall N_{ah}+1 \leq n \leq N$.\\
Due to channel uncertainty, the minimum distance between the constellation points is reduced.
In order to calculate the new minimum distance, the channel uncertainty is approximated using
standard deviations $\vartheta_n$ of the channels of the active ad hoc nodes. As an example, consider
two constellation points $Q_1$ and $Q_2$, which pertain to states $\textbf{s}_1[m] = [1, 0, 1, 0]$ and $\textbf{s}_2[m] =[0, 1, 1, 0]$, respectively. Assuming a Euclidean distance $d[m]$ between $Q_1$ and $Q_2$ under perfect CSI condition and $N_{ah} = 3$, we can approximate the modified Euclidean distance $d_{\mathrm{mod}}$ between them under imperfect CSI condition as 
\begin{equation}
d_{\mathrm{mod}} = \max\{d[m] - 4\vartheta_1 - 4\vartheta_2, 0\},
\end{equation}
where the standard deviations $\vartheta_1$ and $\vartheta_2$ are multiplied by 4 in order to guarantee that this procedure is valid in $99.994\%$ of cases\footnote{Such a high precision is motivated by the reliability condition.} based on the underlying Gaussian distribution of $\hat{h}_n$.
Note, that the channel uncertainty related to node 3 is not taken into account, since this node
is considered active with respect to both constellation points, such that a possible deviation of
channel $\hat{h}_3$ from its expected value $\bar{h}_n$ would affect both points in the same way. Hence, both
points would be shifted in the same direction and the distance between them would remain
unchanged. Through this, the observed symbol constellation might deviate from the expected
symbol constellation. However, it would still be possible to reliably distinguish between the
constellation points for $d_{\mathrm{mod}} > 0$.\\
Apparently, only the channel variations need to be taken into account, which are indicated
by the outcome of a logical XOR operation applied to the respective states pertaining to the
neighboring constellation points. In the example above, $\textbf{s}_1 \bigoplus \textbf{s}_2 = [1, 1, 0, 0]$, such that only the
first two standard deviations $\vartheta_1$ and $\vartheta_2$ are taken into account. Obviously, the distance between constellation points is heavily affected by the channel uncertainty
and might quickly reduce to very small values\footnote{By directly subtracting the weighted standard deviations from the original distance $d[m]$, even negative values can result. In
order to avoid this, we introduce a clipping to zero in (12), since even in this case the harvested power is zero.} even for moderate channel variances $\vartheta_n^2$.
\subsection{Computational complexity}
\label{sec_complexity}
In general, the proposed method does not require highly complex calculations. The most
computationally expensive parts of the algorithm are the state prediction according to Alg. 1
and the calculation of the minimum distance between any two points of the resulting joint
constellation according to Alg. 2 line 6 and Alg. 3 line 11 for SBP and BBP, respectively. However, the distance calculation can be realized via complex-valued
summation, which has a very low computational complexity.
For the state prediction according to Alg. 1, we have $N$ real-valued multiplications per state (see
line 21). Correspondingly, for the SBP we obtain $O(2^NN)$ multiplications for
all $2^N$ states. For the BBP, the number of executions of the Alg. 1 is related
to the number of states pre-stored in $\textbf{u}_j$, which changes from symbol to symbol of the same
block. In the worst case scenario, the number of states in $\textbf{u}_j$ is equal to the total number of
possible states $2^N$. Hence, the resulting number of multiplications based on state prediction for
the BBP is upper bounded by $O(D2^{2N}N)$.
For a more realistic complexity estimation, we determine the average number of active nodes
as $N_{av} =\sum_n\frac{p_nL_n}{p_nL_n+(1-p_n)}$. As an example, we set $p_n = 10^{-2}$ and $L_n = 20,\:\forall n$. Hence, we obtain $N_{av}\approx 0.168N$. Correspondingly, the complexity of SBP and BBP in this case is $O(0.168N2^{0.168N})$ and $O(0.168N2^{0.168N}D)$, respectively. Interestingly,
with increasing probability of transmission or packet length, the average number of observed
nodes and the complexity increase as well. This indicates a non-trivial trade-off between the
length of the duty cycle related to $p_n$ and the computational complexity, which should be taken
into account in the system design.\\
A relatively high computational complexity of the proposed schemes may lead to further power consumption at the relay and substantially reduce the harvested energy. Sometimes, this may render the proposed method even less energy-efficient than the baseline scheme. However, the calculation of the PSF can be done remotely, e.g. at the base station. The required communication overhead from the relay to the base station may solely consist of channel coefficients, since all other parameters, e.g. packet length, modulation type, etc. are typically known to the base station. Of course, in case of mobility of the nodes, the CSI needs to be updated at the base station, which requires additional overhead. However, in the considered scenario, we assume a limited mobility, such that the update frequency is low. In the opposite direction, i.e. from the base station to the relay, the PSF values for all possible state transitions need to be transmitted. In total, there are up to $3^N$ PSF values to be stored at the relay depending on the probability of transmission. Hence, the relay would be able to select the stored PSF value, which corresponds to the current state of the Markov process.\\
If the relay has to calculate the PSF autonomously, the respective computational complexity needs to be reduced e.g. by introducing a maximum PSF, for which the energy harvesting would still be possible. Hence, the algorithm would stop earlier upon reaching this threshold and set PSF to 1. Alternatively, the PSF prediction can be done via machine learning, which would be trained for various channel conditions, states of Markov process and transition probabilities. In this case, the training can still be performed at the base station, such that only the trained predictor would be stored at the relay. Correspondingly, the communication overhead and high computational complexity can be avoided altogether. As a distinct advantage, channel variations would not affect the prediction performance. However, this approach is beyond the scope of this work.
\section{Numerical Results}
\label{sec_4}
In our simulations, we assume that the nodes are randomly deployed with a distance between
3 m and 10 m around the relay according to Fig. 1. Also, an equal (maximum) transmit power $P_t = 20$ dBm for each node, a bandwidth of 100 kHz, and a carrier frequency of 900 MHz are
assumed. For the signal propagation, a Rician flat fading channel with the line-of-sight factor 3,
a path loss exponent 2, and additive white Gaussian noise with respective variances $\sigma^2 = -110$
dBm and $\delta^2 = -75$ dBm are used. For the energy conversion efficiency, we set $\eta$ = 0.5. Each
node transmits a packet of equal length $L_n=L= 20,\: \forall n$ symbols with equal probability $p_n = p,\: \forall n$, where $p$ is a design parameter. For a better accuracy of simulation results, we average over
the outcome of 5000 scenarios for each simulation point. In each scenario, a sequence of 1000
symbols is observed, which results from the overlapping transmissions from all $N$ nodes.\\
In this work, we focus on BPSK transmissions, although higher-order modulations are possible, too, as mentioned earlier. However, with increasing modulation order, the number of constellation points increases, which leads to shorter distances between them and therefore lower harvested energy.
\subsection{Ultrareliability}
Although the main focus of this work is on the maximization of the harvested power, we provide some insight in the expected symbol error rates (SER) and packet error rates (PER). Since the worst-case signal quality with respect to the closest constellation points is $\mathrm{SNR}_\mathrm{mod}[m] \geq 13$dB $\triangleq 20$, the expected SER can be estimated (cf. \cite{goldsmith2005cambridge}) as $\mathrm{SER}\leq Q(\sqrt{2\cdot 20})\approx10^{-10}$, where $Q(\cdot)$ is the complementary Gaussian error integral. Correspondingly, we obtain in case of uncoded transmission $\mathrm{PER}\leq 1-(1-\mathrm{SER})^L\approx \mathrm{SER}\cdot L=10^{-10}L$, where $L$ is the packet length. With coded transmission, such a low symbol error probability leads to nearly zero packet errors after the FEC decoder, which is typical for ultrareliable communication. As a result, the FEC coding rate can be selected very high in order to avoid unnecessary data rate losses. Alternatively, the minimum required $\mathrm{SNR}_{\mathrm{mod}}$ can be reduced, such that the expected PER would meet the PER requirements. Through this, smaller distances between constellation points can be tolerated by the detector, such that more energy can be harvested. In addition, more nodes can be incorporated into the RMA with this strategy without causing any performance degradation. The optimal value for $\mathrm{SNR}_{\mathrm{mod}}$ depends on the employed FEC code, such that a trade-off between coding rate (spectral efficiency), harvested energy and scalability of the proposed method (number of nodes) arises. This trade-off requires a thorough investigation and is beyond the scope of this work.
\subsection{Symbol-based predictor (SBP)}
The average harvested power using a SBP for different numbers of nodes is
shown in Fig. \ref{fig5}. 
\ifCLASSOPTIONdraftcls
\begin{figure}
\centering
\includegraphics[width=0.7\textwidth]{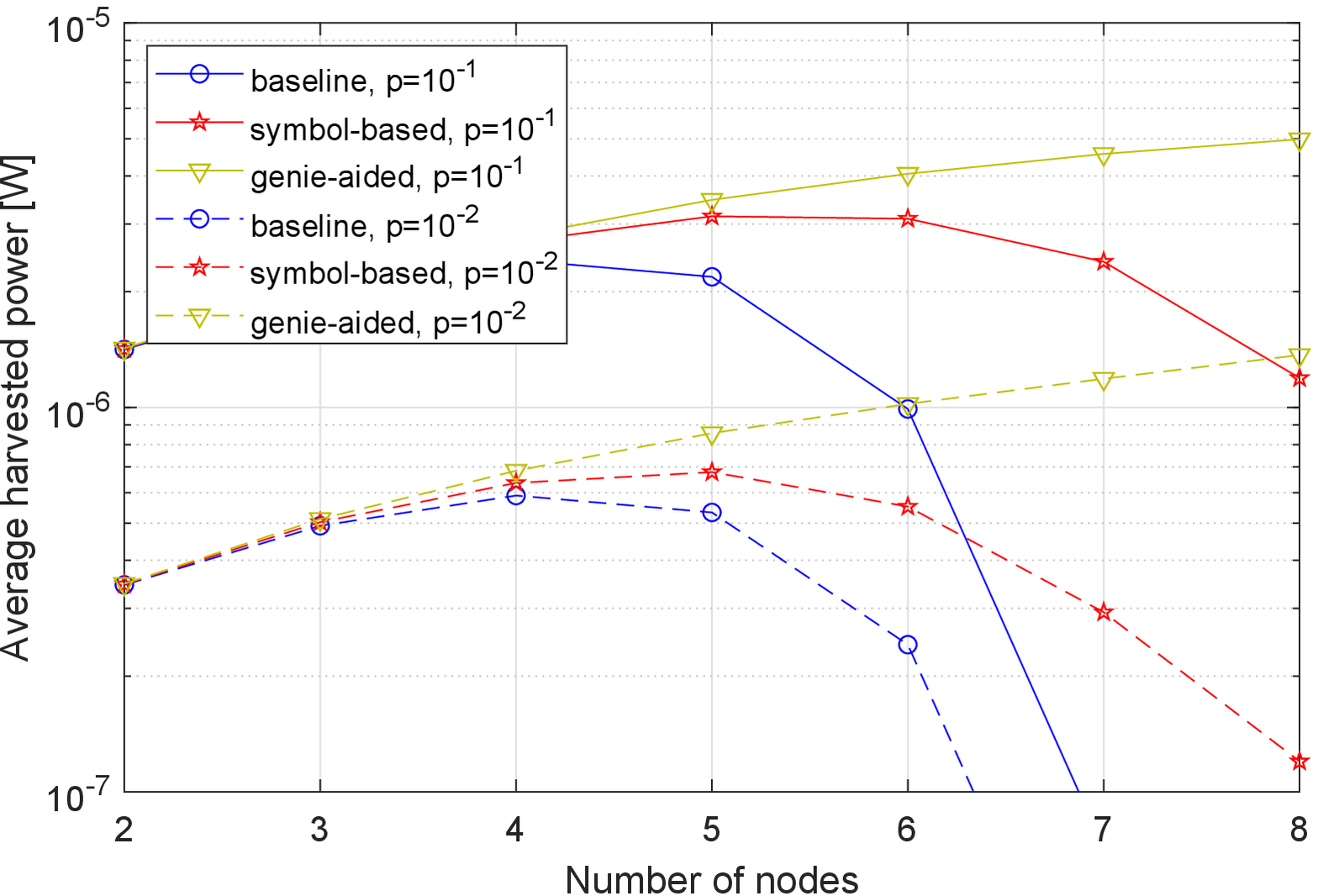}
\caption{Average harvested power for various numbers of nodes and $p\in\{10^{-2}, 10^{-1}\}$.}
\label{fig5}
\end{figure}
\else
\begin{figure}
\includegraphics[width=0.49\textwidth]{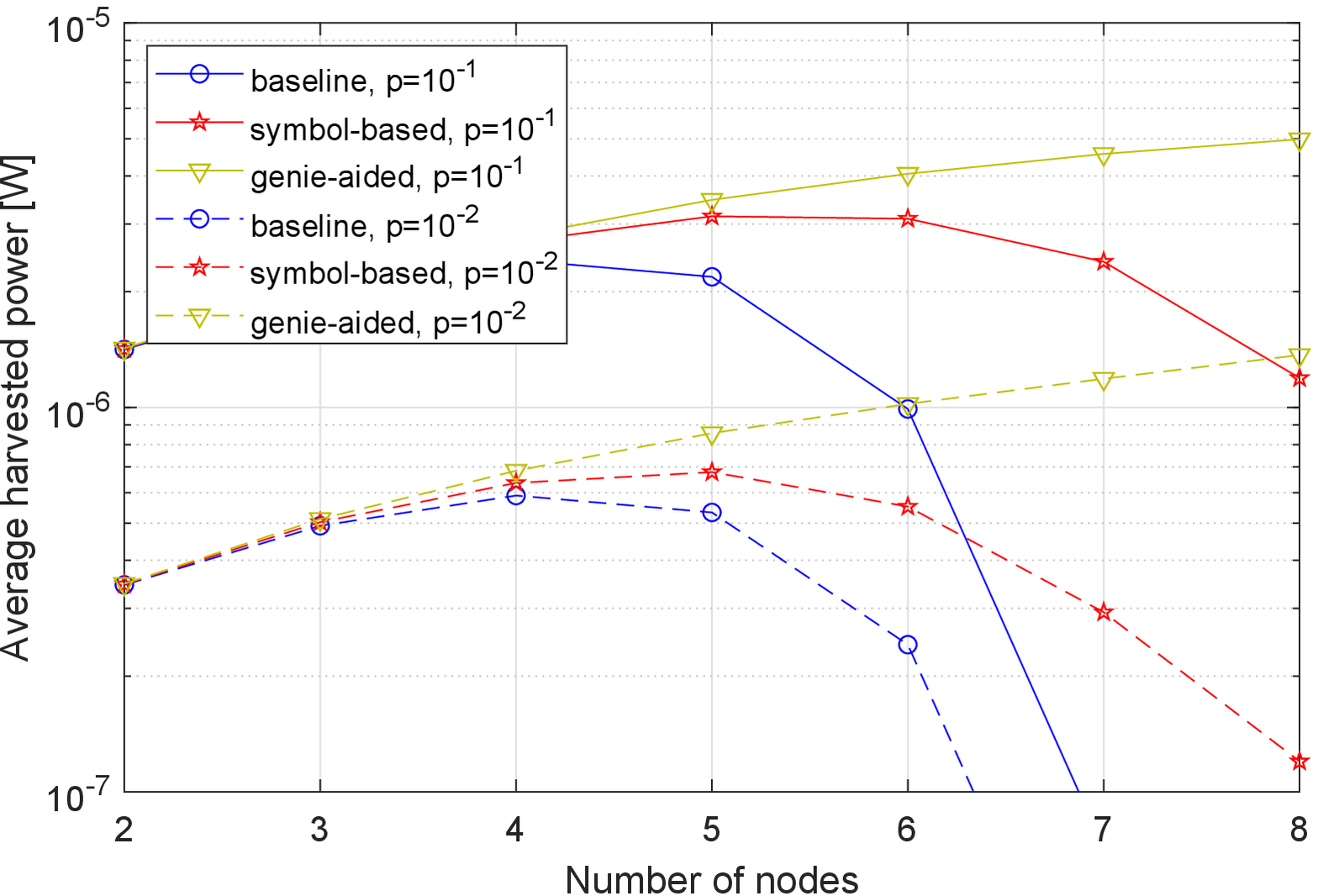}
\caption{Average harvested power for various numbers of nodes and $p\in\{10^{-2}, 10^{-1}\}$.}
\label{fig5}
\end{figure}
\fi
In general, the harvested power increases with the probability of transmission,
since more power is transmitted by the nodes. The baseline scheme has its maximum with
$N=4$ independently of $p$ followed by a steep decrease. This decrease results from the
increasing number of points in the joint constellation, such that the minimum distance between
the points reduces. Correspondingly, less energy can be harvested. A similar behavior is observed with the proposed method. However, the maximum is located around $N=5$ followed by a
slight decrease for $N=6$, such that a gain of 5 dB and 7.5 dB can be observed compared
to the baseline scheme for the respective probabilities of transmission $p \in \{10^{-2}, 10^{-1}\}$. For
$N \geq 7$, the baseline scheme does not allow any reasonable energy harvesting, such that
$P_{\mathrm{harv}} \approx 0$. For $N = 8$ and $p = 0.1$, the proposed method is still capable of providing
$P_{\mathrm{harv}} \approx 1.2$ $\mu$W. In contrast, the genie-aided scheme can provide up to $4.7$ $\mu$W under the same settings, which is 8 dB better than the proposed scheme. The gap between the genie-aided scheme
and the proposed scheme can only be reduced, if the required SNR (which is currently set to
13 dB) for information detection is reduced. However, no reliable communication would be
guaranteed in this case.\\
In Fig. \ref{fig6}, the results for the harvested power are depicted as a function of $p$ for $N = 6$.
\ifCLASSOPTIONdraftcls
\begin{figure}
\centering
\includegraphics[width=0.7\textwidth]{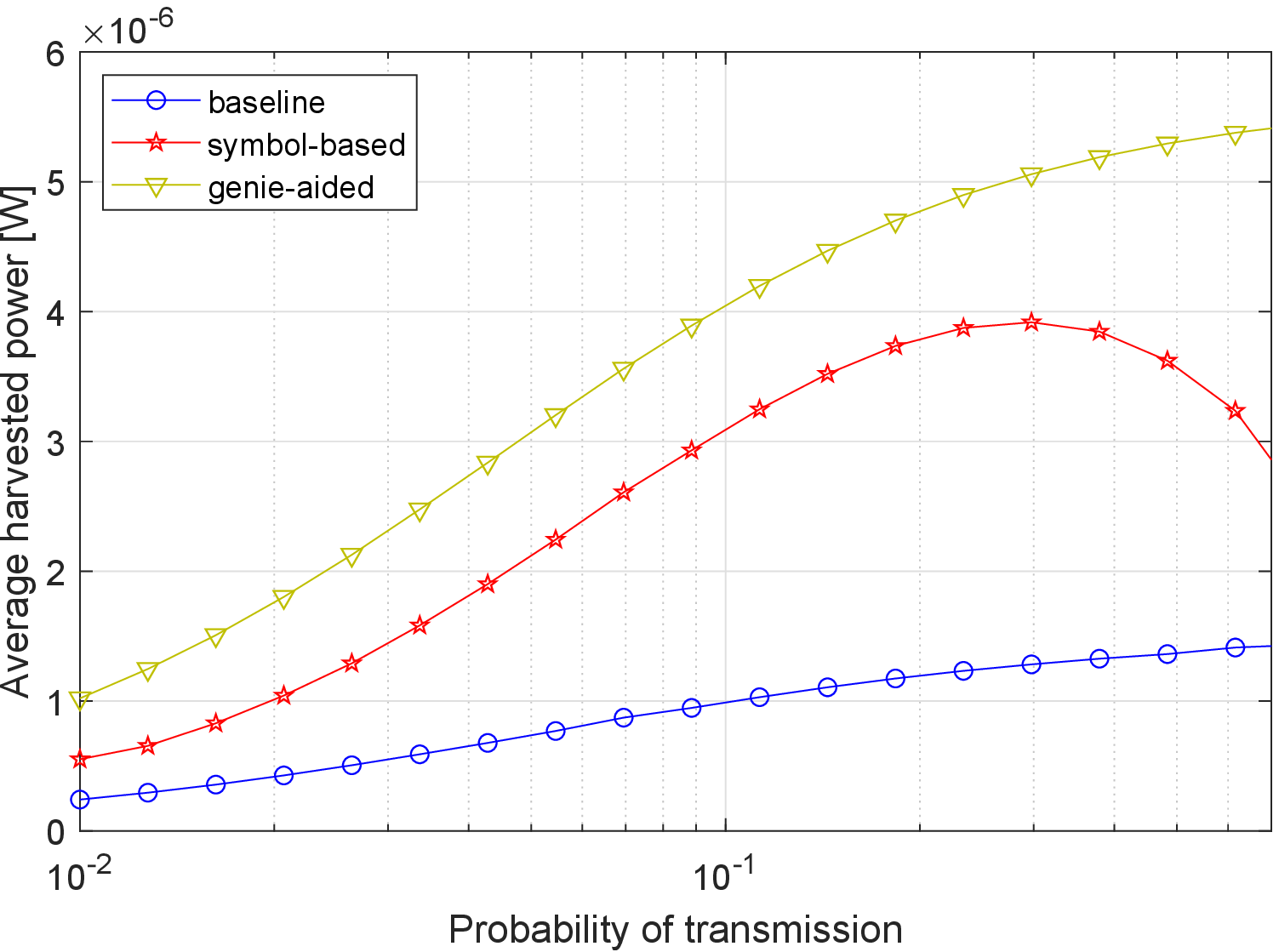}
\caption{Average harvested power for 6 nodes and various probabilities $p$.}
\label{fig6}
\end{figure}
\else
\begin{figure}
\includegraphics[width=0.48\textwidth]{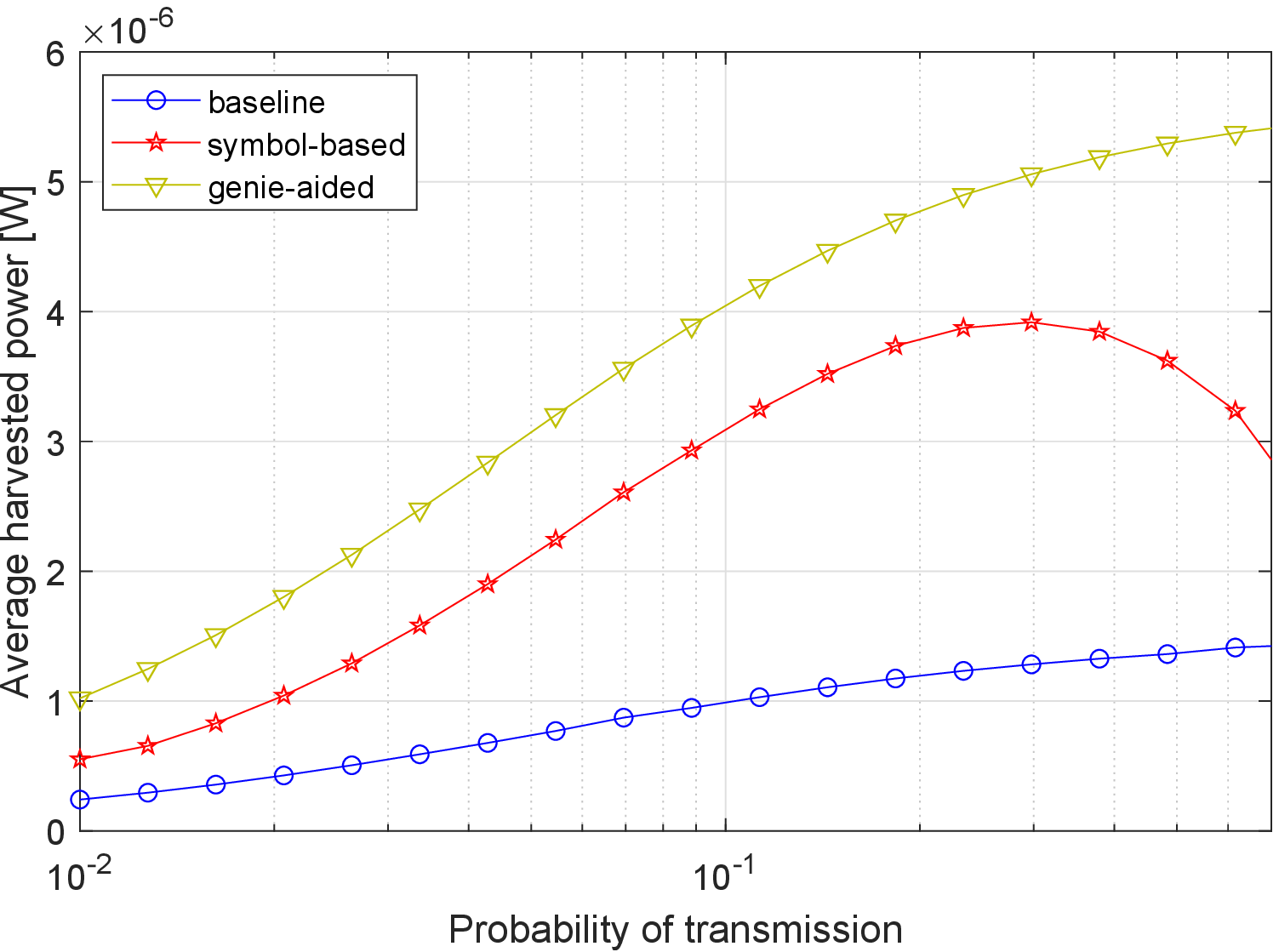}
\caption{Average harvested power for 6 nodes and various probabilities $p$.}
\label{fig6}
\end{figure}
\fi
We observe that both performance bounds increase with increasing probability of transmission,
while the proposed method has a maximum at $p \approx 0.3$. The increase of the harvested power
with increasing $p$ is due to the increased number of packets that are transmitted on average,
such that more energy is also consumed by the nodes according to \eqref{eq:cons} and correspondingly
received at the relay. For $p > 0.3$, the average harvested power decreases, since the nodes are more frequently active and interfere with each other more often. Hence, the state, which pertains
to a small $d_0[m]$, is more likely to occur. This behavior can be typically observed in case of
saturation of the transmit power, i.e. for $P_{\mathrm{consumed},n} > 0.9P_t,\: \forall n$, where the nodes are not able to provide substantially more power and the interference becomes the limiting factor for the
system performance. This situation occurs e.g. for $p > 0.3$ and $L = 20$.\\
In order to show the impact of the packet length, we simulate the SWIPT for $p = 0.01$ and $N = 6$ and various $L$. The results are depicted in Fig. \ref{fig7}. 
\ifCLASSOPTIONdraftcls
\begin{figure}
\centering
\includegraphics[width=0.7\textwidth]{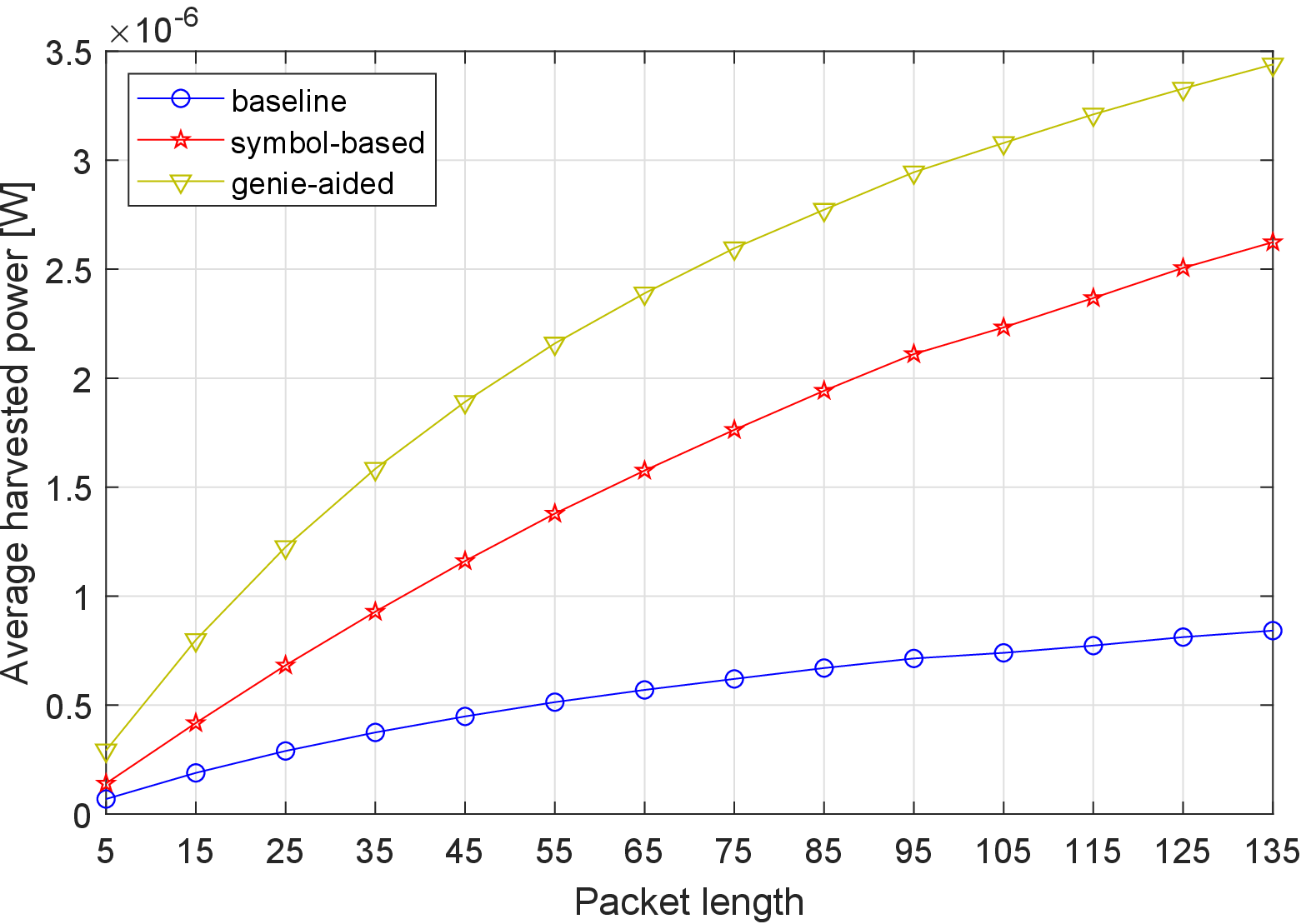}
\caption{Average harvested power for 6 nodes, $p = 0.01$ and packet lengths.}
\label{fig7}
\end{figure}
\else
\begin{figure}
\includegraphics[width=0.49\textwidth]{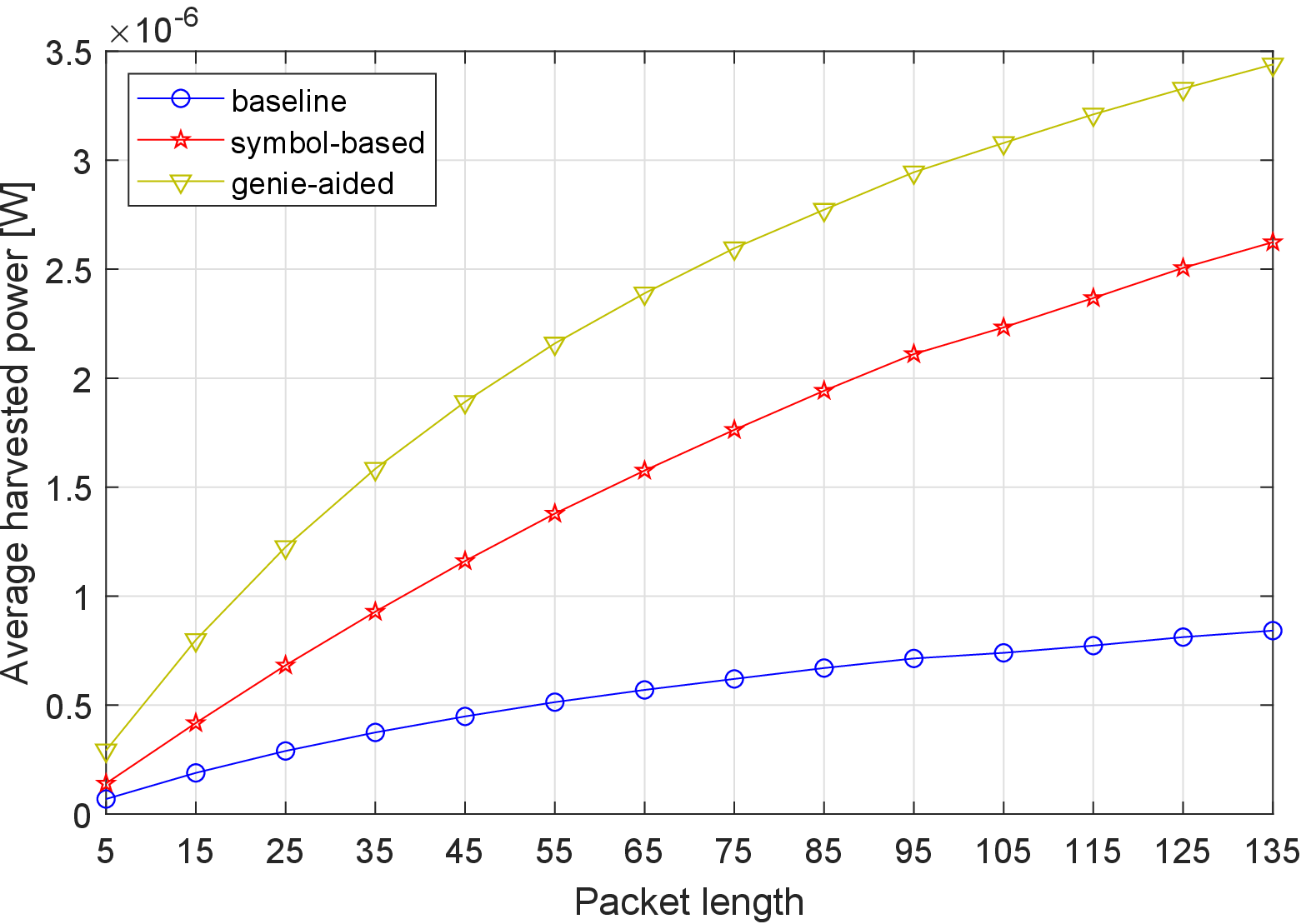}
\caption{Average harvested power for 6 nodes, $p = 0.01$ and packet lengths.}
\label{fig7}
\end{figure}
\fi
The average harvested power increases monotonically
with increasing $L$ for all three schemes, since the nodes transmit more symbols on average,
as can be deduced from (1). Surprisingly, there is no maximum for the proposed scheme as
compared to Fig. 6. As mentioned earlier, the performance is dominated by interference for
$P_{\mathrm{consumed},n} > 0.9P_t,\:\forall n$, which occurs with $L > 900$, if we assume $p = 0.01$. Hence, we can deduce from Fig. 7, that it is beneficial to make the packets longer, since more power can be
harvested using the proposed practical method. However, long packets are not reasonable in the considered scenario, since the flexibility of RMA associated with short packet transmission decreases with increasing packet length. Hence, a trade-off between harvested power and packet length will be considered in future system design. Furthermore, we observe a gap of $\approx 4.8$ dB between the proposed scheme and the baseline for $L = 105$ in Fig. 7, whereas the gap between the genie-aided and the proposed scheme is only 1.25 dB, which is very promising.\\ 
Note, that the behavior of the average harvested power may change in case of non-linear efficiency of the energy harvesting circuits. Typically, each active node contributes to the joint symbol constellation with additional symbol points and thus reduces the distance between the symbols. Hence, the signal power at the input of the energy harvesting circuits reduces as well. As known from the previous works on energy harvesting (cf. e.g. \cite{boshkovska2015practical}), the efficiency of non-linear energy harvesters is usually very low in case of low input power. Correspondingly, less power can be harvested, if the nodes remain in the active state for a longer time, i.e. if the packet length or the transmission probability increases.
\subsection{Block-based predictor (BBP)}
For the BBP, we simulate the SWIPT for $p = 0.1$ and different numbers of
nodes. The length of the packets is set to $L = 20$. Furthermore, the number of packets is set to
10000 for a better accuracy. The results are shown in Fig. \ref{fig8}. 
\ifCLASSOPTIONdraftcls
\begin{figure}
\centering
\includegraphics[width=0.7\textwidth]{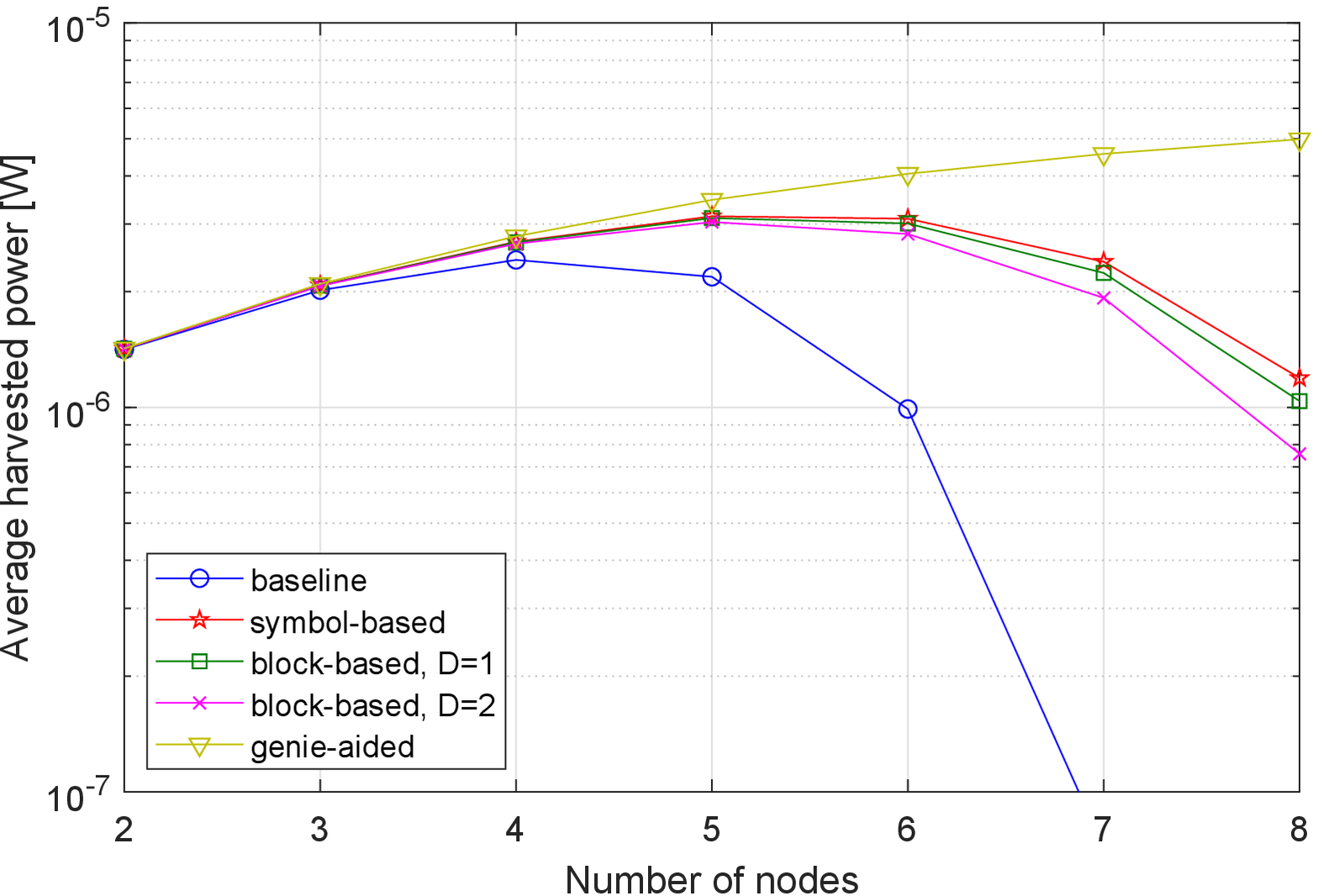}
\caption{Average harvested power for various numbers of nodes and $p = 0.1$.}
\label{fig8}
\end{figure}
\else
\begin{figure}
\includegraphics[width=0.48\textwidth]{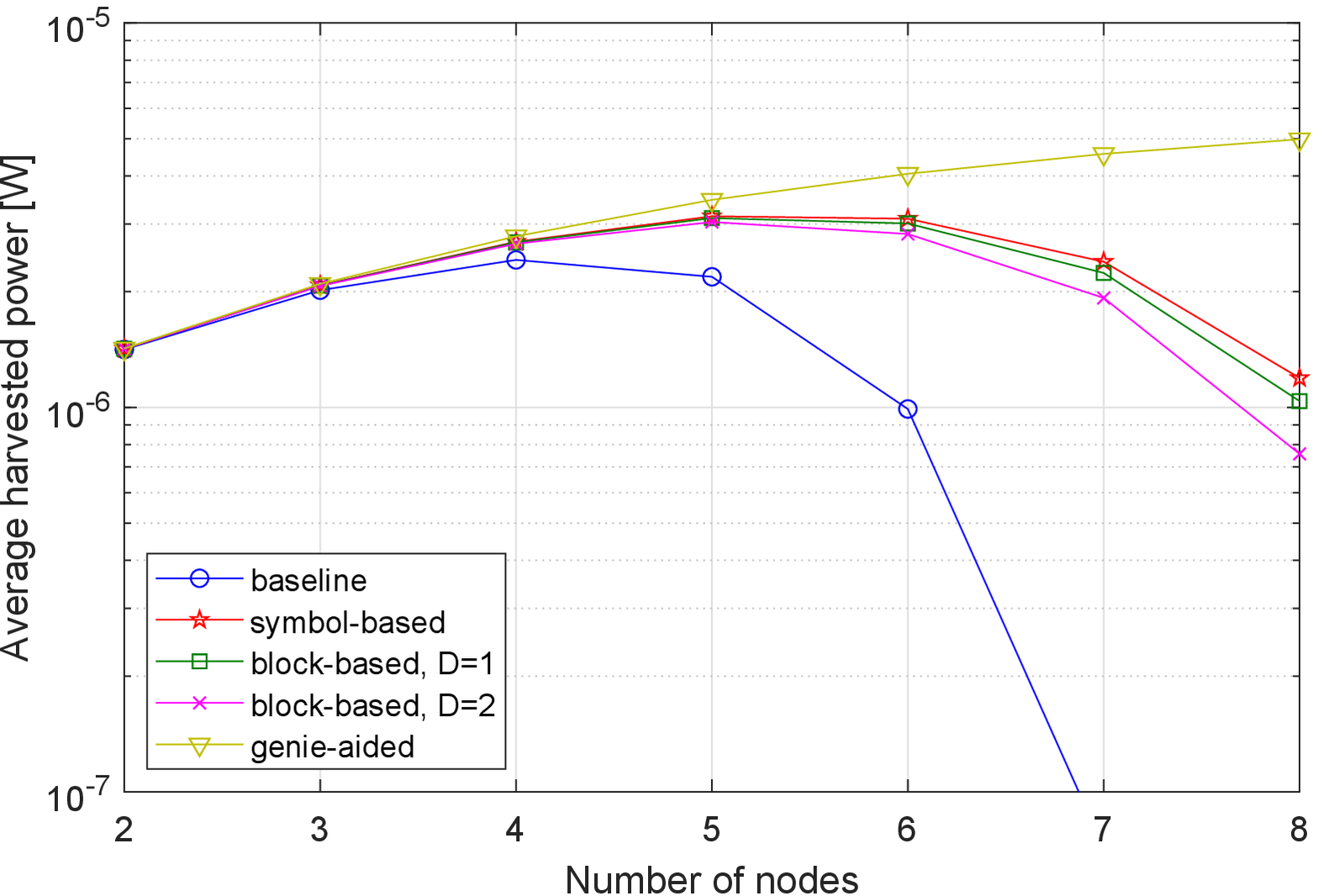}
\caption{Average harvested power for various numbers of nodes and $p = 0.1$.}
\label{fig8}
\end{figure}
\fi
As we can see, the performance of
the BBP is very similar to that of the SBP. However, with
increasing delay $D$, the average harvested power decreases more and more, especially with a
large number of adjacent transmissions. Interestingly, the gap between these schemes increases
with increasing number of nodes as well, which is due to the increasing number of possible states,
which can occur within the window of $2D$ symbols. The corresponding joint constellation has
therefore more points, such that the minimum distance between them decreases, which leads to
a lower harvested power, as discussed earlier. Surprisingly, the performance degradation due to a longer prediction interval is not large, such that the BBP is a valid practical
solution for the considered problem. However, the complexity of this predictor is much higher
than for the SBP, which might restrict the choice of $D$ to only a few symbols.
In our simulations, we also observed that the relative performance gap between the SBP
and the BBP remains approximately constant independently of packet length
and probability of transmission. The reason for this behavior is that the accuracy of BBP is related to the number of states, which can be observed within the block. This
number of states is mainly dictated by the number of nodes as long as $L \gg D$.
\subsection{Imperfect CSI}
In order to investigate the performance of the predictor in case of imperfect CSI, we assume
that all links have the same channel uncertainty with respect to the transmission channels, i.e. $\vartheta_n^2=\alpha|\bar{h}|^2,\:\forall n$, where $\alpha$ is the uncertainty factor. The results for $N = \{4, 6\}$, $L = 20$, $p = 0.1$ are shown in Fig. \ref{fig9}. 
\ifCLASSOPTIONdraftcls
\begin{figure}
\centering
\includegraphics[width=0.7\textwidth]{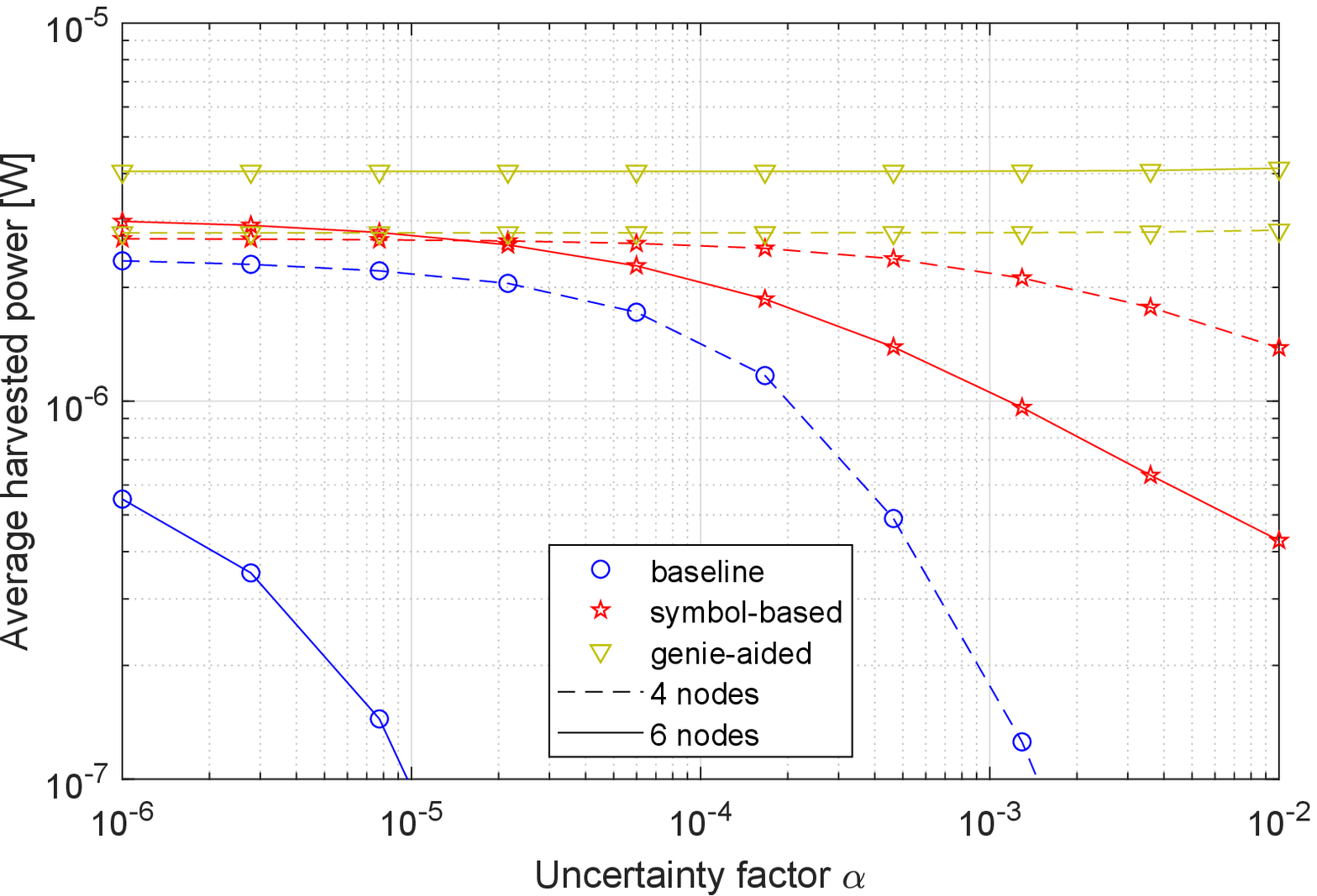}
\caption{Average harvested power vs. channel variance.}
\label{fig9}
\end{figure}
\else
\begin{figure}
\includegraphics[width=0.49\textwidth]{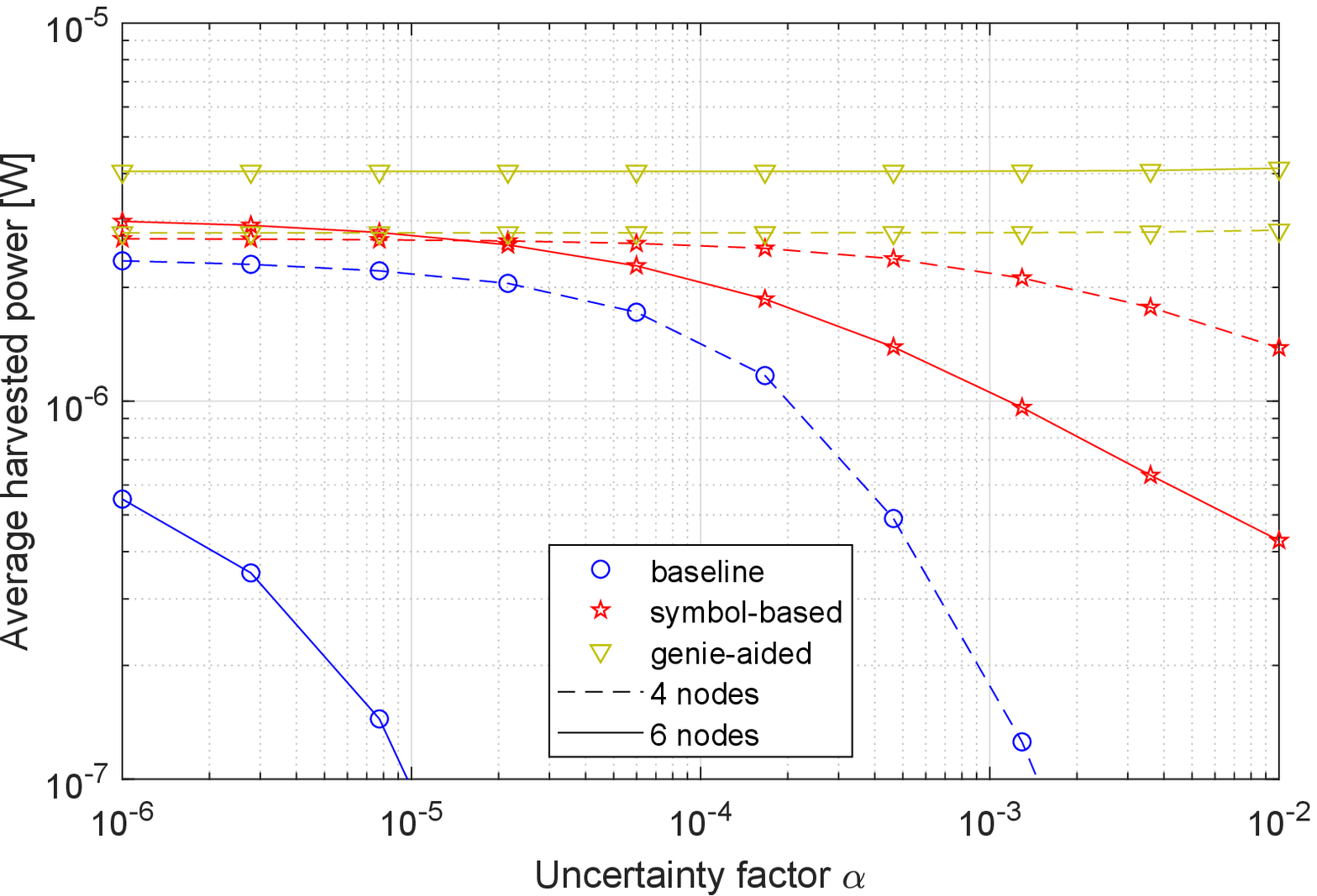}
\caption{Average harvested power vs. channel variance.}
\label{fig9}
\end{figure}
\fi
Similarly to Fig. \ref{fig5}, we observe that the average harvested power using the baseline scheme and the proposed solution decrease with increasing number of nodes,
whereas the genie-aided performance bound increases. Furthermore, we observe an increasing
performance degradation for both practical schemes with increasing channel uncertainty given by factor $\alpha$. In particular, almost no power can be harvested using the baseline scheme and
$N = 4$, if $\alpha > 10^{-3}$. On the contrary, the proposed method is less vulnerable to imperfect
CSI, such that the respective harvested power decreases much slower. Correspondingly, with
$N = 4$ and $\alpha = 10^{-2}$ approximately $50\%$ of the genie-aided power can be harvested using
the SBP. Unfortunately, this gap increases substantially with increasing $N$,
such that only $10\%$ of the genie-aided power can be harvested with the proposed method in case
of 6 nodes and $\alpha = 10^{-2}$. However, the channel uncertainty $\alpha$ is usually much lower in practice
due to the typically stationary deployment and correspondingly a very thorough synchronization
and channel acquisition.
\section{Conclusion}
\label{sec_5}
In this paper, we investigated the design of dynamically adjustable power splitting at a relay
device for the randomly scheduled short packet transmissions and reliable communication. Since the number of interfering packets in each symbol interval is unknown before the
power splitting, the optimal splitting factor is predicted based on the previously received symbols.
This has been done while guaranteeing reliable communication in terms of extremely low
symbol and packet error rate.\\ 
We proposed two methods, symbol-based and block-based predictors, respectively,
which exploit the knowledge of the packet length and the probability of transmission by
each node. The symbol-based predictor calculates the optimal power splitting factor for only one
symbol interval ahead without taking into account possible delays due to high computational
complexity. In contrast, the block-based predictor calculates the optimal splitting factor for a
block of symbols of a given length. Both methods have shown a substantial gain of the harvested
power compared to the baseline scheme, where no prediction is done. On the other hand, a
significant gap between the proposed methods and the theoretical upper bound can be observed,
which can only be bridged by sacrificing the reliability of symbol detection. In future work, the limits of the proposed scheme under ultrareliability condition will be investigated.\\ 
In addition, we observed that the optimal number of nodes, for which the proposed methods are especially beneficial, is relatively low, i.e. between 5 and 8. In order to accommodate more nodes, either the system requirements need to be relaxed or an alternative hybrid medium access should be employed, such that the considered RMA would be part of a larger OMA protocol. In this case, hundreds of nodes can be accommodated. Unlike traditional OMA, each orthogonal medium resource block would be occupied by multiple nodes that transmit randomly. The corresponding harvested power would comprise the contributions from all resource blocks. \\
Furthermore, the impact of imperfect channel state information on the prediction performance has been addressed
and the corresponding performance degradation has been reduced via adaptation of the splitting
factor to the expected possible variations of the communication channels.\\ For a deeper insight into the predictor design and in particular the trade-off between complexity and accuracy of detection, we provide a complexity analysis for both proposed methods. This analysis is important for the future development and implementation of ultrareliable uplink communication.
\section*{Appendix}
\begin{algorithm}[H]
\caption{Selection of relevant states}
\begin{algorithmic}[1]
\REQUIRE $p_n$, $L_n$, $1\leq n\leq N$, $\tilde{\textbf{s}}[m-l],\:\forall 2\leq l\leq L_n$, $\textbf{s}_k[m]$, $\textbf{g}_m$, $k$
\ENSURE $g_m$
\STATE Obtain $s_n[m-1],\:\forall n$ from $\tilde{\textbf{s}}[m-1]$ and $s_n[m],\:\forall n$ from $\tilde{\textbf{s}}[m]$;
\FOR{$n\leftarrow 1$ \TO $N$}
	\IF{$s_n[m-1]=0\: \cap\: s_n[m]=0$}
		\STATE $\mathrm{Pr}(s_n[m]\:|\: s_n[m-1])\leftarrow 1-p_n$;
	\ELSIF{$s_n[m-1]=0\: \cap\: s_n[m]=1$}
		\STATE $\mathrm{Pr}(s_n[m]\:|\: s_n[m-1])\leftarrow p_n$;
	\ELSIF{$s_n[m-1]=1\: \cap\: s_n[m]=0$}
		\IF{$s_n[m-l]=1,\: 2\leq l\leq L_n$}
			\STATE $\mathrm{Pr}(s_n[m]\:|\: s_n[m-1])\leftarrow 1-p_n$;
		\ELSE
			\STATE $\mathrm{Pr}(s_n[m]\:|\: s_n[m-1])\leftarrow 0$;
		\ENDIF
	\ELSE
		\IF{$s_n[m-l]=0,\: l<L_n$}
			\STATE $\mathrm{Pr}(s_n[m]\:|\: s_n[m-1])\leftarrow 1$;
		\ELSE
			\STATE $\mathrm{Pr}(s_n[m]\:|\: s_n[m-1])\leftarrow p_n$;
		\ENDIF
	\ENDIF
\ENDFOR
\STATE $\mathrm{Pr}(\textbf{s}_k[m])\leftarrow \displaystyle\prod_n\mathrm{Pr}(s_n[m]\:|\: s_n[m-1])\cdot 1$
\IF{$\mathrm{Pr}(\textbf{s}_k[m])>10^{-8}$}
	\STATE Determine constellation points which pertain to state $\textbf{s}_k[m]$;
	\STATE Append the constellation points to $\textbf{g}_m$.
\ENDIF
\end{algorithmic}
\end{algorithm}

\begin{algorithm}[H]
\caption{Symbol-based prediction}
\begin{algorithmic}[1]
\REQUIRE $p_n$, $L_n$, $1\leq n\leq N$, $\tilde{\textbf{s}}[m-l],\:\forall 2\leq l\leq L_n$, 
\ENSURE $\rho[m]\leftarrow \min\Big\{\max\Big\{\frac{20\delta^2}{(0.5d_0[m])^2-20\sigma^2}, 10^{-2}\Big\}, 1\Big\}$
\STATE Generate $2^N$ possible states $\textbf{s}_k[m]$;
\STATE Initialize storage $\textbf{g}_m$;
\FOR{$k\leftarrow 1$ \TO $2^N$}
	\STATE Execute Alg. 1;
\ENDFOR
\STATE Determine the minimum Euclidean distance $d_0[m]$ between any two points stored in $\textbf{g}_m$.
\end{algorithmic}
\end{algorithm}

\begin{algorithm}[H]
\caption{Block-based prediction}
\begin{algorithmic}[1]
\REQUIRE $p_n$, $L_n$, $1\leq n\leq N$, $\textbf{u}_{m-l}$, $\tilde{\textbf{s}}[m-l],\:\forall 2\leq l\leq L_n$
\ENSURE $\rho[q(m)]\leftarrow \min\Big\{\max\Big\{\frac{20\delta^2}{(0.5d_0[m])^2-20\sigma^2}, 10^{-2}\Big\}, 1\Big\}$, $\textbf{u}_m$
\STATE Generate $2^N$ possible states $\textbf{s}_k[m],\: q(m)\leq m < q(m)+D$;
\STATE Initialize storage $\textbf{g}_m,\: q(m)\leq m < q(m)+D$ for points and $\textbf{u}_m, \: q(m)\leq m < q(m)+D$ for states;
\STATE Execute Alg. 2 lines 1-5;
\STATE Store $\textbf{s}_k[q(m)]$ for which $\mathrm{Pr}(\textbf{s}_k[q(m)])>10^{-8}$ in $\textbf{u}_{q(m)}$ (remove duplicates);
\FOR{$j\leftarrow q(m)+1$ \TO $q(m)+D$}
	\FORALL{$\textbf{s}_k[j-1]$ from $\textbf{u}_{j-1}$}
		\STATE Set $\tilde{\textbf{s}}[j-1]\leftarrow \textbf{s}_k[j-1]$;
		\STATE Proceed as in Alg. 2 lines 1-5;
		\STATE Store $\textbf{s}_k[j]$ for which $\mathrm{Pr}(\textbf{s}_k[j]\:|\:\textbf{s}_k[j-1])\mathrm{Pr}(\textbf{s}_k[j-1])>10^{-8}$ in $\textbf{u}_j$ (remove duplicates);
	\ENDFOR
	\STATE Determine the minimum Euclidean distance $d_0[j]$ between any two points stored in $\textbf{g}_m$.
\ENDFOR
\STATE Choose the minimum Euclidean distance $d_0[q(m)]=\min_jd_0[j]$.
\end{algorithmic}
\end{algorithm}
\bibliographystyle{IEEEtran}
\bibliography{Literature}

\begin{thebibliography}{10}
\providecommand{\url}[1]{#1}
\csname url@samestyle\endcsname
\providecommand{\newblock}{\relax}
\providecommand{\bibinfo}[2]{#2}
\providecommand{\BIBentrySTDinterwordspacing}{\spaceskip=0pt\relax}
\providecommand{\BIBentryALTinterwordstretchfactor}{4}
\providecommand{\BIBentryALTinterwordspacing}{\spaceskip=\fontdimen2\font plus
\BIBentryALTinterwordstretchfactor\fontdimen3\font minus
  \fontdimen4\font\relax}
\providecommand{\BIBforeignlanguage}[2]{{%
\expandafter\ifx\csname l@#1\endcsname\relax
\typeout{** WARNING: IEEEtran.bst: No hyphenation pattern has been}%
\typeout{** loaded for the language `#1'. Using the pattern for}%
\typeout{** the default language instead.}%
\else
\language=\csname l@#1\endcsname
\fi
#2}}
\providecommand{\BIBdecl}{\relax}
\BIBdecl

\bibitem{kisseleff2019ultrareliable}
S.~Kisseleff, S.~Chatzinotas, and B.~Ottersten, ``{Ultrareliable SWIPT using
  Unscheduled Short Packet Transmissions},'' in \emph{Proc. of IEEE ICC,
  B5G-URLLC Workshop}, 2019.

\bibitem{atzori2010the}
L.~Atzori, A.~Iera, and G.~Morabito, ``The internet of things: {A} survey,''
  \emph{{Computer} networks}, vol.~54, no.~15, pp. 2787--2805, 2010.

\bibitem{raza2017low}
U.~Raza, P.~Kulkarni, and M.~Sooriyabandara, ``Low power wide area networks: An
  overview,'' \emph{IEEE Communications Surveys \& Tutorials}, vol.~19, no.~2,
  pp. 855--873, 2017.

\bibitem{saito2013nonorthogonal}
Y.~Saito, Y.~Kishiyama, A.~Benjebbour, T.~Nakamura, A.~Li, and K.~Higuchi,
  ``Non-orthogonal multiple access ({NOMA}) for cellular future radio access,''
  in \emph{Proc. of IEEE Vehicular Technology Conference (VTC Spring)}, 2013,
  pp. 1--5.

\bibitem{lieske2016decoding}
H.~Lieske, G.~Kilian, M.~Breiling, S.~Rauh, J.~Robert, and A.~Heuberger,
  ``Decoding performance in low-power wide area networks with packet
  collisions,'' \emph{{IEEE Transactions on Wireless Communications}}, vol.~15,
  no.~12, pp. 8195--8208, 2016.

\bibitem{kisseleff2018optimal}
S.~Kisseleff, J.~Kneissl, G.~Kilian, and W.~Gerstacker, ``Optimal {MAP}
  detection in presence of burst interference for low power wide area
  networks.''\hskip 1em plus 0.5em minus 0.4em\relax Proc. of IEEE Global
  Communications Conference, 2018, pp. 1--5.

\bibitem{mengali1997synchronization}
U.~Mengali and A.~D'Andrea, \emph{Synchronization Techniques for Digital
  Receivers}.\hskip 1em plus 0.5em minus 0.4em\relax Plenum Press, {NY}, 1997.

\bibitem{polyanskiy2010channel}
Y.~Polyanskiy, H.~Poor, and S.~Verd'u, ``Channel coding rate in the finite
  blocklength regime,'' \emph{{IEEE Transactions on Information Theory}},
  vol.~56, no.~5, p. 2307, 2010.

\bibitem{durisi2016shortpacket}
G.~Durisi, T.~Koch, J.~¨Ostman, Y.~Polyanskiy, and W.~Yang, ``Short-packet
  communications over multiple-antenna rayleighfading channels,'' \emph{{IEEE
  Transactions on Communications}}, vol.~64, no.~2, pp. 618--629, 2016.

\bibitem{durisi2016toward}
G.~Durisi, T.~Koch, and P.~Popovski, ``Toward massive, ultrareliable, and
  low-latency wireless communication with short packets,'' \emph{Proceedings of
  the {IEEE}}, vol. 104, no.~9, pp. 1711--1726, 2016.

\bibitem{popovski2018wireless}
P.~Popovski, J.~Nielsen, C.~Stefanovic, E.~Carvalho, E.~Strom,
  K.~Trillingsgaard, A.~Bana, D.~Kim, R.~Kotaba, J.~Park, and R.~Sorensen,
  ``Wireless access for ultra-reliable low-latency communication: {Principles}
  and building blocks,'' \emph{{IEEE Network}}, vol.~32, no.~2, pp. 16--23,
  March 2018.

\bibitem{she2017radio}
C.~She, C.~Yang, and T.~Quek, ``Radio resource management for ultra-reliable
  and low-latency communications,'' \emph{{IEEE Communications Magazine}},
  vol.~55, no.~6, pp. 72--78, June 2017.

\bibitem{mousaei2017optimizing}
M.~Mousaei and B.~Smida, ``Optimizing pilot overhead for ultra-reliable
  short-packet transmission.''\hskip 1em plus 0.5em minus 0.4em\relax Proc. of
  IEEE ICC, May 2017, pp. 1--5.

\bibitem{she2018joint}
C.~She, C.~Yang, and T.~Quek, ``Joint uplink and downlink resource
  configuration for ultra-reliable and low-latency communications,''
  \emph{{IEEE Transactions on Communications}}, vol.~66, no.~5, pp. 2266--2280,
  May 2018.

\bibitem{singh2018contentionbased}
B.~Singh, O.~Tirkkonen, Z.~Li, and M.~Uusitalo, ``Contention-based access for
  ultra-reliable low latency uplink transmissions,'' \emph{{IEEE Wireless
  Communications Letters}}, vol.~7, no.~2, pp. 182--185, April 2018.

\bibitem{hu2018relayingenabled}
Y.~Hu, M.~Gursoy, and A.~Schmeink, ``{Relaying-Enabled Ultra-Reliable
  Low-Latency Communications in 5G},'' \emph{{IEEE} Network}, vol.~32, no.~2,
  pp. 62--68, March 2018.

\bibitem{gu2018ultrareliable}
Y.~Gu, H.~Chen, Y.~Li, and B.~Vucetic, ``Ultra-reliable short-packet
  communications: {Half}-duplex or full-duplex relaying?'' \emph{{IEEE Wireless
  Communications Letters}}, vol.~7, no.~3, pp. 348--351, June 2018.

\bibitem{vu2017ultrareliable}
T.~Vu, C.~Liu, M.~Bennis, M.~Debbah, M.~Latva-aho, and C.~Hong,
  ``{Ultra-Reliable and Low Latency Communication in {mmWave}-Enabled Massive
  MIMO Networks},'' \emph{{IEEE Communications Letters}}, vol.~21, no.~9, pp.
  2041--2044, 2017.

\bibitem{souza2018ultrareliable}
O.~L.~A. L{\'o}pez, E.~M.~G. Fern{\'a}ndez, R.~D. Souza, and H.~Alves,
  ``Ultra-reliable cooperative short-packet communications with wireless energy
  transfer,'' \emph{IEEE Sensors Journal}, vol.~18, no.~5, pp. 2161--2177,
  2018.

\bibitem{shi2014joint}
Q.~Shi, L.~Liu, W.~Xu, and R.~Zhang, ``Joint transmit beamforming and receive
  power splitting for {MISO SWIPT} systems,'' \emph{{IEEE} Transactions on
  Wireless Communications}, vol.~13, no.~6, pp. 3269--3280, June 2014.

\bibitem{kodheli2017integration}
O.~Kodheli, A.~Guidotti, and A.~Vanelli-Coralli, ``Integration of satellites in
  5{G} through {LEO} constellations,'' in \emph{Proc. of {IEEE Global
  Communications Conference}}, December 2017, pp. 1--6.

\bibitem{akyildiz2010wireless}
I.~Akyildiz and M.~Vuran, \emph{Wireless sensor networks}.\hskip 1em plus 0.5em
  minus 0.4em\relax John Wiley \& Sons, 2010, vol.~4.

\bibitem{6525243}
G.~{Kilian}, H.~{Petkov}, R.~{Psiuk}, H.~{Lieske}, F.~{Beer}, J.~{Robert}, and
  A.~{Heuberger}, ``Improved coverage for low-power telemetry systems using
  telegram splitting,'' in \emph{European Conference on Smart Objects, Systems
  and Technologies (Smart SysTech)}, June 2013, pp. 1--6.

\bibitem{1532220}
S.~Cui, A.~Goldsmith, and A.~Bahai, ``Energy-constrained modulation
  optimization,'' \emph{IEEE Transactions on Wireless Communications}, vol.~4,
  no.~5, pp. 2349--2360, Sep. 2005.

\bibitem{zhang2013mimo}
R.~Zhang and C.~Ho, ``{MIMO broadcasting for simultaneous wireless information
  and power transfer},'' \emph{{IEEE Transactions on Wireless Communications}},
  vol.~12, no.~5, pp. 1989--2001, 2013.

\bibitem{8330063}
M.~{Rajabi}, N.~{Pan}, S.~{Claessens}, S.~{Pollin}, and D.~{Schreurs},
  ``{Modulation Techniques for Simultaneous Wireless Information and Power
  Transfer With an Integrated Rectifier–Receiver},'' \emph{IEEE Transactions
  on Microwave Theory and Techniques}, vol.~66, no.~5, pp. 2373--2385, May
  2018.

\bibitem{claessens2019multitone}
S.~Claessens, N.~Pan, D.~Schreurs, and S.~Pollin, ``{Multitone FSK modulation
  for SWIPT},'' \emph{IEEE Transactions on Microwave Theory and Techniques},
  vol.~67, no.~5, pp. 1665--1674, 2019.

\bibitem{boshkovska2015practical}
E.~Boshkovska, D.~Ng, N.~Zlatanov, and R.~Schober, ``{Practical non-linear
  energy harvesting model and resource allocation for {SWIPT} systems},''
  \emph{{IEEE Communications Letters}}, vol.~19, no.~12, pp. 2082--2085, 2015.

\bibitem{clerckx2018beneficial}
B.~Clerckx and J.~Kim, ``On the beneficial roles of fading and transmit
  diversity in wireless power transfer with nonlinear energy harvesting,''
  \emph{IEEE Transactions on Wireless Communications}, vol.~17, no.~11, pp.
  7731--7743, 2018.

\bibitem{liu2013chua}
L.~Liu, R.~Zhang, and K.-C. Chua, ``{Wireless} information and power transfer:
  {A} dynamic power splitting approach,'' \emph{{IEEE Transactions on
  Communications}}, vol.~61, no.~9, pp. 3990--4001, 2013.

\bibitem{verdu1998cambridge}
S.~Verdu, \emph{Multiuser Detection}.\hskip 1em plus 0.5em minus 0.4em\relax
  Cambridge University Press, 1998.

\bibitem{goldsmith2005cambridge}
A.~Goldsmith, \emph{Wireless Communications}.\hskip 1em plus 0.5em minus
  0.4em\relax Cambridge University Press, 2005.

\bibitem{meyr1997digital}
H.~Meyr, M.~Moeneclaey, and S.~Fechtel, \emph{Digital communication receivers:
  synchronization, channel estimation, and signal processing}.\hskip 1em plus
  0.5em minus 0.4em\relax John Wiley \& Sons, Inc., 1997.

\end{thebibliography}

\end{document}